\documentclass[aps,prd,preprint,floatfix,nofootinbib]{revtex4-1}
\usepackage{graphicx}
\usepackage{epic}
\usepackage{eepic}
\usepackage{latexsym}
\usepackage{amssymb,amsmath}

\newcommand{\eq}[1]{(\ref{#1})}
\newcommand{\be}{\begin{equation}}
\newcommand{\ee}{\end{equation}}
\newcommand{\bea}{\begin{eqnarray}}
\newcommand{\eea}{\end{eqnarray}}

\newcommand{\hs}[1]{\hspace{#1 mm}}

\newcommand{\df}{\dot{\phi}}

\newcommand{\vc}{\vec{k}}
\newcommand{\vx}{\vec{x}}
\newcommand{\vv}{|0\right>}
\newcommand{\vvc}{\left<0|}
\newcommand{\tvv}{|\Omega\right>}
\newcommand{\tvvc}{\left<\Omega|}
\newcommand{\tk}{\tilde{k}}

\newcommand{\ak}{a_{\vec{k}}}
\newcommand{\akd}{a_{\vec{k}}^\dagger}

\def\b{\beta}

\def\d{\delta}
\def\D{\Delta}
\def\e{\epsilon}

\def\f{\phi}
\def\fr{\frac}

\def\l{\lambda}

\def\m{\mu}
\def\n{\nu}

\def\s{\sigma}

\def\th{\theta}

\def\O{\Omega}

\def\o{\omega}

\def\del{\partial}

\let\bm=\bibitem
\def\nn{\nonumber}

\begin{document}

\title{On {\Large $i\epsilon$} Prescription in Cosmology} 

\author{Ali Kaya}
\email[]{ali.kaya@boun.edu.tr}
\affiliation{Bo\~{g}azi\c{c}i University, Department of Physics, 34342, Bebek, \.{I}stanbul, Turkey}

\date{\today}

\begin{abstract}
This is a technical note on the $i\epsilon$ prescription in cosmology where we consider a self-interacting scalar field in the Poincare patch of the de Sitter space whose Hamiltonian has explicit time dependence. We use both path integral and operator formalisms to work out the evolution of states from asymptotic past infinity with $i\epsilon$ prescription, which becomes nontrivial even in the free theory, and explicitly show how arbitrary states are projected onto the vacuum. We establish that in perturbation theory the $i\e$ prescription can be implemented in Weinberg's commutator formula by just inserting $\epsilon$ dependent convergence factors that make the oscillating time integrals at infinity meaningful. 
\end{abstract}

\maketitle

\section{Introduction}

Quantum field theory has an intricate mathematical structure and the $i\e$ prescription is a peculiar technicality. One immediately encounters it in the quantization of {\it free} fields in flat space in obtaining the momentum space Feynman propagator. In the interacting theory, the $i\e$ prescription amounts to choosing the proper time integration contour, which projects the free vacuum onto the full interacting vacuum (for an operator formalism derivation see \cite{book1} and for the path integral approach see \cite{book2}). 

Not surprisingly, the $i\e$ prescription also appears for quantum fields in cosmological Friedmann-Robertson-Walker (FRW) backgrounds. For example, it arises when expressing the free field de Sitter propagator in terms of the coordinate length function in position space (see e.g. \cite{i1,i2,i3,i4}). In the presence of interactions, it is also implemented like in the flat space field theory by picking up a proper time integration contour including a small imaginary piece, which is supposed to project onto the full interacting vacuum of the theory (see e.g. \cite{c1,c2,c3,c4,c5,c6}, see also \cite{c7} for an alternative approach). 

However, there are crucial differences between flat space and FRW quantum field theories (QFTs) like the explicit time dependence of the Hamiltonian and the absence of a globally defined invariant vacuum state. Therefore it is not clear at all why giving a small imaginary component to the time variable still selects the vacuum state in the asymptotic past. Indeed, the time independence of the Hamiltonian and consequently its eigenvectors are crucial in establishing the projection in flat space (see e.g. \cite{book1}). Despite this fact, the projection in the cosmological setting is either taken for granted without proof or one repeats the flat space argument as if the Hamiltonian is time independent  (see e.g. \cite{baum}), presumably by assuming some form of adiabaticity at early times. More crucially, one should consider the normalization of the resulting state vector after the projection since it is a non-unitary operation and in the flat space calculation this gives the cancellation of vacuum to vacuum graphs in Green functions \cite{book1}. This issue seems to be completely ignored in the cosmological context.  

In \cite{wein1}, Weinberg gives a neat formula that expresses a full Heisenberg picture operator in terms of nested commutators of the corresponding interaction picture operator and the interaction Hamiltonian, which is very suitable for in-in perturbation theory. Due to the asymmetry of contour specification in the $i\e$ prescription, this operator identity is generally broken down when it is sandwiched between states to read the vacuum expectation values. Although there can be ways of incorporating the $i\e$ prescription while still preserving the commutator structure as discussed in \cite{c3}, one may hesitate to use Weinberg's formula in cosmological perturbation theory because of this discrepancy. An aim of this work is to clarify this issue. 

Because the $i\e$ prescription is commonly identified with selecting the true interacting vacuum state, one may wonder why it ever appears in the free theory where the vacuum is exactly solvable. Below, we will show that the generic time evolution dictated by the Schr\"{o}dinger equation starting from (or extending to) infinite past (or future) becomes technically problematic since there appear infinite ambiguous phases. To see this in an elementary example, just take the quantum mechanical harmonic oscillator where the superposition of the vacuum and the first exited state is chosen with equal weights as the initial state at $t_0$. The corresponding wave-function can be determined as $\left.|\psi\right>= e^{-iE_0(t-t_0)}\left.\vv+e^{-iE_1(t-t_0)}a^\dagger\left.\vv$ and the expectation value of the position operator $\left<\psi|x|\psi\right>\propto \cos[(E_0-E_1)(t-t_0)]$ becomes ill defined in the limit $t_0\to-\infty$. As we will see, the $i\e$ prescription enables one to take such limits arising in cosmology meaningfully. 

In this paper we would like to formalize the $i\e$ prescription for a self-interacting scalar field defined in the Poincare patch of the de Sitter space. Namely, we would like to give a derivation of the standard $i\e$ prescription used in cosmology from first principles by using both path integral and operator formalisms. Although it is very well known and somehow elementary, we first extensively discuss the vacuum of the free theory by carefully paying attention to the past asymptotic infinity limit. As we will show, this essentially elucidates the emergence of $i\e$ terms in cosmological loop calculations. 

\section{The Problem}

Let $\f$ be a self-interacting minimally coupled real scalar field in a fixed FRW background that has the scale factor $a(t)$. The action can be written as
\be
S=\int d^3 x\,dt\,\, a^3\left[\fr12 \dot{\f}^2-\fr{1}{2a^2} (\del_i\f)^2-V(\f)\right],
\ee
where dot denotes the time derivative and $V(\f)$ is the interaction potential possibly including a mass term $\fr12 m^2\f^2$. The corresponding Hamiltonian is given by
\be
H=\int d^3x \left[\fr{1}{2a^3}P_\f^2+\fr{1}{2}\,a\, (\del_i\f)^2+a^3V(\f)\right],\label{ham}
\ee
where $P_\f=a^3\df$. In this paper we consider the cosmological patch of the de Sitter space as the background and thus take $a=a_0\,e^{H t}$ when the scale factor is explicitly needed.

\subsection{Free Theory}

The canonical quantization of the {\it free theory} can be achieved straightforwardly. Introducing the ladder operators and the mode functions as
\be\label{fa}
\f(t,\vx)=\int \fr{d^3 k}{(2\pi)^{3/2}}\left[ e^{i\vc.\vx}\f_k(t)\,\ak+e^{-i\vc.\vx}\f^*_k(t)\,\akd\right],
\ee
the canonical commutation relation
\be
[\f(t,\vx),P_\f(t,\vx\,')]=i\d^3(\vx-\vx\,')
\ee
can be satisfied by imposing $[\ak,a_{\tk}^\dagger]=\d^3(k-\tk)$ and the Wronskian condition
\be
\f_k\df_k^*-\df_k\f_k^*=\fr{i}{a^3}.\label{w}
\ee
The (Heisenberg picture) field equations imply the mode equation
\be\label{me}
\ddot{\f}_k+3H\df_k+\left(m^2+\fr{k^2}{a^2}\right)\f_k=0
\ee 
and one may define the corresponding {ground state} as usual by 
\be\label{fv}
\ak\left.\vv=0.
\ee
This completes the quantization of the free theory while there is still an arbitrariness for the choice of the mode function $\f_k$, which reflects a well known feature i.e. the non-uniqueness of vacuum in a cosmological background.   

Nevertheless, the ground state should be defined as {\it the eigenstate of the Hamiltonian with the lowest eigenvalue}. This is not just because the Hamiltonian is related to the energy of the system, but even further it dictates time evolution. The free Hamiltonian \eq{ham} is quadratic in field variables, i.e. the system is actually an infinite dimensional harmonic oscillator which has a unique ground state at any given time. In the function space representation, where $P_\f(\vx)=-i\d/d\f(\vx)$, the {\it instantaneous} ground state at $t_0$ can be found as a Gaussian wave-functional
\be\label{v1}
\Psi_0[\f,t_0]=C\exp\left[-\fr12 \int d^3k\,a^3\o(k,t_0)\,\f(\vc)\,\f(-\vc)\right]
\ee
where $C$ is a normalization constant, 
\be
\o(k,t)=\left[m^2+\fr{k^2}{a(t)^2}\right]^{1/2}
\ee
and 
\be\label{ft}
\f(\vx)=\int\fr{d^3k}{(2\pi)^{3/2}}\,e^{i\vc.\vx}\,\f(\vc).
\ee
On the other hand, the vacuum defined at $t_0$ will become excited at a later time by the Schr\"{o}dinger equation due to explicit time dependence in the Hamiltonian, which can identified as particle creation. 

In the operator description that uses the Heisenberg picture, one can use \eq{fa} in the free Hamiltonian to obtain
\bea
H=&&\fr12 a^3\int d^3k\left[|\df_k|^2+\o(k,t)^2|\f_k|^2\right]\left(\ak\akd+\akd\ak\right)\label{haa}\\
&&+\left[\df_k^2+\o(k,t)^2\f_k^2\right]\ak\, a_{-\vec{k}}+\left[(\df^*_k)^2+\o(k,t)^2(\f^*_k)^2\right]\akd\, a^\dagger_{-\vec{k}}.\nn
\eea
The first line is the standard harmonic oscillator Hamiltonian and $\left.\vv$ becomes the instantaneous ground state {\it if and only if} the second line of \eq{haa} vanishes at $t_0$. Together with the Wronskian condition \eq{w}, this implies (up to a seemingly unimportant phase)  
\be\label{inc}
\f_k(t_0)=\fr{1}{a(t_0)^{3/2}\sqrt{2\o(k,t_0)}},\hs{5}\df_k(t_0)=\fr{-i}{a(t_0)^{3/2}}\sqrt{\fr{\o(k,t_0)}{2}}.
\ee
Therefore, when the mode equation \eq{me} is solved with the above initial conditions, the state $\left.\vv$, which obeys $\ak\left.\vv=0$, becomes the instantaneous vacuum at time $t_0$. In the function space representation, $\left.\vv$ appears as in \eq{v1}, i.e. $\left<\f\vv=\Psi_0[\f,t_0]$. 

Let us now focus on the de Sitter space and the Bunch-Davies vacuum whose mode function is given by
\be\label{bdm}
\f^{BD}_{k}=\left(\fr{\pi}{4H\,a^3}\right)^{1/2}\,e^{i\pi\n/2}\,H^{(1)}_\n\left(\fr{k}{aH}\right),
\ee
where $\n=\sqrt{9/4-m^2/H^2}$ and $H^{(1)}_\n$ is the Hankel function of first kind. This must correspond to a vacuum at {\it past asymptotic infinity} where all (relevant) modes become subhorizon. We would like to confront the Bunch-Davies vacuum to \eq{inc}. Using the asymptotic form of the Hankel function one may find that
\be
\lim_{a\to0}\f_k^{BD}\,=\,e^{ik/(aH)}\fr{1}{a\sqrt{2k}}\left[1+\fr{i(2H^2-m^2)}{2Hk}a+{\cal O}(a^2)\right].
\ee
Comparing this to \eq{inc} in the $t_0\to-\infty$ limit one sees that only the leading order terms match and the remaining higher order terms differ with each other. Likewise, the Hamiltonian \eq{haa} evaluated for the Bunch-Davies mode function \eq{bdm} in the early time limit gives 
\bea
\lim_{a\to0}H=&&\fr12 \int d^3k\left[\fr{k}{a}+{\cal O}(a^0)\right]\left(\ak\akd+\akd\ak\right)\label{hbd}\\
&&+e^{2ik/(Ha)}\left[H+{\cal O}(a)\right]\ak\, a_{-\vec{k}}+e^{-2ik/(Ha)}\left[H+{\cal O}(a)\right]\akd\, a^\dagger_{-\vec{k}},\nn
\eea
which shows that in fact $\left.\vv$ is {\it not} an eigenstate\footnote{One may insist that in the limit $a\to0$ the first line in \eq{hbd}, which has order $a^{-1}$, becomes much larger than the second line, which has order $a^0$. However, the second line cannot simply be neglected as compared to the first because different operators are involved. One may also tempt to include a small decaying piece to the mode functions to kill the oscillating second line. Indeed, such a behavior arises after a careful treatment of the $i\e$ prescription, see \eq{fep} below.} of the Hamiltonian since  the second line above survives the limit. 

To find the correct ground state one should respect \eq{inc} and to satisfy it at some $t_0$ one must take the superposition of positive and negative frequency solutions. One then hopes that only one solution survives in the limit $t_0\to-\infty$, which would subsequently yield the standard Bunch-Davies vacuum. This will turn out to be the case up to an infinite phase, which is mostly harmless. To make the discussion simple, let us consider the massless scalar which has the following Bunch-Davies mode function 
\be\label{bdmsl}
f_{BD}=\fr{1}{a\sqrt{2k}}\exp[ik/(Ha)]\left[1+\fr{iHa}{k}\right].
\ee
Eq. \eq{inc} implies that the actual mode function corresponding to the ground state at $t_0$ must be chosen as
\be\label{fc}
\f_k=c_1 f_{BD}+c_2f_{BD}^*
\ee
where
\be\label{cc1}
c_1=\left[1-\fr{iHa(t_0)}{2k}\right]\exp[-ik/Ha(t_0)],\hs{5}c_2=\fr{-iHa(t_0)}{2k}\exp[ik/Ha(t_0)].
\ee
Clearly, $a(t_0)\to0$ gives $c_2\to0$ and $c_1\to \exp[-ik/Ha(t_0)]$, and  one recovers the Bunch-Davies mode function \eq{bdmsl} up to the {\it divergent phase factor} that sits in $c_1$. This infinite phase factor is harmless and actually cancels out when one considers the vacuum expectation values like $\vvc\f(t,\vx)\f(t,\vx\,')\vv$. Yet, it might be significant when superposition of states are involved in the expectation values. Consider for instance $\left<g\right|\f(t,\vc_1)\f(t,\vc_2)\left| g\right>$, where $\left.|g\right>=\int d^3 k\, g(\vc)\,\akd\left.\vv$ and $g(\vc)$ is an arbitrary function obeying $\int d^3k |g|^2=1$ that gives  $\left<g|g\right>=1$. The Fourier transformed field $\f(t,\vc)$ is defined as in \eq{ft} and one finds that
\bea
\left<g\right|\f(t,\vc_1)\f(t,\vc_2)\left|g\right>=&&|\f_{k_1}|^2\d^3(\vc_1+\vc_2)\nn\\
&&+g(\vc_1)g^*(-\vc_2)\f_{k_1}(t)\f^*_{k_2}(t)+g(\vc_2)g^*(-\vc_1)\f_{k_2}(t)\f^*_{k_1}(t). \label{gg}
\eea
Now, using \eq{fc} in \eq{gg} gives the phase $\exp[i(k_1-k_2)/Ha(t_0)]$ in the second line which makes the expression ill defined in the $t_0\to-\infty$ limit. If, on the other hand, one recalls that the correlation function \eq{gg} is a {\it distribution} in the momentum space and it should be interpreted only after smearing out with test functions, then in that case $t_0\to-\infty$ limit completely kills the second line following smearing. This final result differs from the Bunch-Davies formula, which can be obtained by using \eq{bdmsl} in \eq{gg}.  

To make the discussion even more intriguing, we could have determined the initial conditions \eq{inc} with specific and perfectly valid $t_0$ dependent phases as  
\be\label{inc2}
\f_k(t_0)=\fr{\exp[ik/Ha(t_0)]}{a(t_0)^{3/2}\sqrt{2\o(k,t_0)}},\hs{5}\df_k(t_0)=\fr{-i\exp[ik/Ha(t_0)]}{a(t_0)^{3/2}}\sqrt{\fr{\o(k,t_0)}{2}}
\ee
so that \eq{cc1} becomes 
\be\label{cc2}
c_1=\left[1-\fr{iHa(t_0)}{2k}\right],\hs{5}c_2=\fr{-iHa(t_0)}{2k}\exp[2ik/Ha(t_0)].
\ee
Here, the phase in $c_1$ goes away, $c_1\to1$ and $c_2\to0$  in the limit $a(t_0)\to0$; as a result all the fuss about infinite phases disappears and \eq{gg} becomes the standard Bunch-Davies result.  
 
The ambiguity is actually related to the temporal asymptotic behavior in quantum theory which becomes intricate because the {\it unitary} time evolution dictated by the Schr\"{o}dinger equation necessarily produces divergent phases when the evolution period is taken to infinity. In the second case above, the problem is pushed to the "initial conditions'' \eq{inc2}, which then gives the Bunch-Davies mode function \eq{bdmsl} in the limit. This explains how the well known Heisenberg picture evolution of free operators in de Sitter space given by the Bunch-Davies mode functions hides the issue; namely it is at the expense of an ambiguity in the past asymptotic infinity. Therefore extra care is needed for the textbook interpretation that the state $\left.\vv$ is actually released from past infinity in the corresponding Schr\"{o}dinger picture. As pointed out below \eq{cc1}, this complication can be ignored in the free theory when calculating the vacuum expectation values which are the main observables of cosmological interest. Nonetheless, it inevitably reappears in the presence of interactions and the $i\e$ prescription offers a way of dealing with it.  

\subsection{Interacting Theory}\label{2b}

Let us begin our discussion of the interacting theory with a review of the interaction picture. Although changing pictures is a standard procedure, some of the definitions will be important later. It is convenient to first analyze the problem in the Schr\"{o}dinger picture. Assume the system is released at time $t_0$ and introduce the following unitary time evolution operators 
\be\label{u}
U_H(t)=T\exp\left[-i\int_{t_0}^t H(\f,P_\f,t')\,dt'\right],\hs{5}U_0(t)=T\exp\left[-i\int_{t_0}^t H_0(\f,P_\f,t')\,dt'\right],
\ee
where $T$ denotes time ordering, $(\f,P_\f)$ are the time independent Schr\"{o}dinger picture operators defined at $t_0$ and $H$, $H_0$ are respectively the full and the free Hamiltonians whose explicit time dependence are indicated in \eq{u}.  For any given (time independent Schr\"{o}dinger picture) operator $O$, one can calculate its expectation value at time $t$ by evolving the initial state from $t_0$ to $t$ by $U_H$. Taking, for the moment, the free vacuum \eq{fv} as the initial state one may define
\be
\left<O\right>\equiv \vvc\right. U_H(t)^\dagger \,O\,U_H(t)\left.\vv.
\ee
To be able to calculate this expression perturbatively, one can introduce the interaction picture quantities as follows 
\be
U_I(t)=U_0(t)^\dagger U_H(t), \hs{5} O_I(t)=U_0(t)^\dagger O U_0(t)
\ee
that give
\be\label{oi}
\left<O\right>=\vvc \right.U_I(t)^\dagger \,O_I(t)\,U_I(t)\left.\vv.
\ee
The interaction picture unitary evolution operator obeys
\be
i\dot{U}_I(t)=H_I\left(\f_I(t),P_{\f I}(t),t\right)\,U_I(t),
\ee
which can be solved as
\be\label{fui}
U_I(t)=T\exp\left[-i\int_{t_0}^t H_I\left(\f_I(t'),P_{\f I}(t'),t'\right)\,dt'\right],
\ee
where $H_I=H-H_0$. As shown in \cite{wein1}, it is possible to expand \eq{oi} order by order in $H_I$ which gives 
\be\label{inp}
\left< O\right>=\sum_{N=0}^{\infty} i^N \int_{t_0}^t dt_N\int_{t_0}^{t_N}dt_{N-1}...\int_{t_0}^{t_2}dt_1\vvc [H_I(t_1),[H_I(t_2),...[H_I(t_N),O(t)]...]\vv,
\ee
where $H_I(t)$ stands for $H_I\left(\f_I(t),P_{\f I}(t),t\right)$ and the zeroth order contribution is $\vvc O_I(t)\vv$. This nice result can be used as the basis for the in-in perturbation theory. 

Everything (up to the issues of renormalization) is well defined in \eq{inp} for finite $t_0$, but $t_0\to-\infty$ limit is problematic as the time integrals become oscillating and non-convergent e.g. for the Bunch-Davies vacuum in de Sitter space (in subsection \ref{3b} we give an explicit example). The infinite phase problem, which has been concealed in the standard treatment of the free theory, inevitably reappears now. It is possible to make these integrals convergent by giving a small imaginary piece to the time variable \cite{c1}, which suggests the following modification of \eq{oi} that has been extensively used in the literature 
\be\label{ue}
\left<O\right>=\vvc\right. U_I^\e(t)^\dagger \,O_I(t)\,U_I^\e(t)\left.\vv,
\ee
where
\be\label{ue2}
U_I^\e(t)=T\exp\left[-i\oint_{t_0}^t H_I(t')dt'\right],\hs{5}t_0=-\infty(1-i\e)
\ee
and the integral is along a contour from $t_0$ to $t$ in the complex plane. Based on the same reasoning used in the flat space QFT, this $i\e$ deformation is justified on the basis that it projects the free vacuum $\left.\vv$ onto the full interacting one $\left.\tvv$ so that
\be\label{tvv}
\tvvc\right. O\left.\tvv(t)=\lim_{\e\to0}\,\vvc\right. U_I^\e(t)^\dagger \,O_I(t)\,U_I^\e(t)\left.\vv,
\ee
where both $\left.\tvv$ and $\left.\vv$ are supposed to be the vacua at past infinity.\footnote{On the other hand, dismissing the bra and the ket vectors in \eq{inp}, the identity still holds as an operator statement relating the Heisenberg and the interaction picture operators; and the convergence issue of the time integrals persists. This shows that the state chosen in calculating the expectation values is not the origin of the problem.} Note that $U_I^\e(t)$ is {\it not a unitary} operator and one may check that the commutator formula \eq{inp} is broken down since the contours of integration are different in $U_I^\e(t)$ and $U_I^\e(t)^\dagger$, which prevent grouping the two terms under a single integral that otherwise produces a commutator. A precise specification of the contour in the complex plane that goes from $t$ to $t_0$ (which also fixes the contour from $t$ to $t_0^*$) is also needed since the integrand depends explicitly (and possibly non-analytically) on time. 

Before discussing why \eq{tvv} is problematic, let us briefly review how the projection occurs {\it when the Hamiltonian has no explicit time dependence.} Let $\left.\tvv$ be the {\it unique} ground state of $H$ with energy $E_\O$ and let $\left.|\psi\right>$ is an arbitrary normalized vector that has some overlap with $\left.\tvv$, i.e. $c_\O\equiv \tvvc\psi\right>\not=0$.  By expanding $\left.|\psi\right>$ in the orthonormal energy eigenbasis as $\left.|\psi\right>=c_\O\left.\tvv+...$ one may see that
\be\label{lh}
e^{-\l H}\left.|\psi\right>=c_\O e^{-\l E_\O}\,\left.\tvv+...
\ee
and therefore 
\be
\lim_{\l\to+\infty}\,\fr{1}{c_\O}\,e^{\l E_\O}e^{-\l H}\left.|\psi\right>=\left.\tvv,
\ee
since $E_\O$ is the minimum energy eigenvalue. Consequently, the following identity holds for any operator $O$ and the state $\left.|\psi\right>$;
\be\label{po}
\tvvc O\tvv=  \lim_{\l\to+\infty}\,\fr{1}{e^{-2\l E_\O}|\tvvc\psi\right>|^2}\left<\psi|\right.e^{-\l H}Oe^{-\l H}\left.|\psi\right>.
\ee
In the calculation of (time ordered) Green functions in the flat space QFT, modifying the time integrals  such as to include a small imaginary component as in \eq{ue2} more or less\footnote{Although the basic method is the same, the actual calculation is a bit more complicated and the projection only works for the free vacuum, see \cite{book1}.} gives the same structure in \eq{po} with  $\l=+\infty$. Moreover, in the same calculation the denominator corresponding to \eq{po} can be identified as the vacuum to vacuum amplitude that normalizes the Green functions \cite{book1}. These justify the use of $i\e$ prescription in the flat space QFT.  

Let us now address why \eq{tvv} is questionable. 

i) Normalization: When $O$ is chosen to be the identity operator, \eq{tvv} should yield $1$ because $\tvvc I \tvv=1$. However, $U_I^\e$ as defined in \eq{ue2} is no longer a unitary operator so this property is not guaranteed at all even after $\e\to0$ (as noted slightly above, the normalization is nontrivial and plays a crucial role in the flat space QFT). Note that the commutator formula \eq{inp} satisfies this property straightforwardly.  

ii) It is not clear how the projection onto the interacting vacuum occurs when the Hamiltonian depends explicitly on time. In principle, the time ordered exponential integral in \eq{ue2} should produce a term like $e^{-\l H}$ in \eq{lh} with diverging $\l$. One can assume adiabaticity at very early times, say before time $t_a$, which is physically viable when all the modes\footnote{In fact, adiabaticity cannot be entirely valid when the whole range of modes $k\in(0,\infty)$ is considered to be physical. However, it is clearly established by many examples that an infrared cutoff is necessary in de Sitter space to make the theory meaningful. With a suitable infrared cutoff, which must be fixed as a comoving scale \cite{br}, adiabaticity becomes a good approximation at early enough times when the cutoff mode is sufficiently subhorizon.} are well inside the horizon. Earlier than $t_a$ the time dependence becomes very weak and one may approximate
\be \label{adi}
T\exp\left[-i\int_{t_0}^{t_a} H_I(t')dt'\right]\simeq \exp\left[-i(t_a-t_0)H_I\right].
\ee
For $t_0=-\infty(1-i\e)$, this produces the required structure for a projection, but {\it not} onto the ground state of the full Hamiltonian $H$ but of $H_I=H-H_0$, which is not the desired result (note that for the self-interacting scalar field $H_I=V-\fr12 m^2\f^2$ and it is not even clear how this projection occurs since $H_I$ has no proper eigenvectors). 

iii) If the projection works properly, it should not matter which state is taken in \eq{tvv} in calculating the vacuum expectation values. Consider, for example, the following normalized vector  
\be\label{01}
\left.|01\right>=c\left.\vv+\int d^3 k\, g(\vc)\,\akd\left.\vv,
\ee
where $|c|^2+\int d^3k |g|^2=1$ implying $\left<01|01\right>=1$. Then one must have 
\be\label{tp}
\tvvc\right. O\left.\tvv(t)=\lim_{\e\to0}\,\vvc\right. U_I^\e(t)^\dagger \,O_I(t)\,U_I^\e(t)\left.\vv=\lim_{\e\to0}\,\left<01|\right.U_I^\e(t)^\dagger \,O_I(t)\,U_I^\e(t)\left.|01\right>. 
\ee
It is not evident how this equation can be satisfied for all arbitrary $c$ and $g(\vc)$ by simply giving a small imaginary piece to the oscillating time integrals. Contrary, one may check that \eq{tp} is not valid in simple examples in in-in perturbation theory. 

To sum up there is a problem of past-eternal time evolution in QFT on cosmological backgrounds. This shows up already in the free theory when one carefully examines the asymptotic behavior of the mode functions. In the presence of  interactions, the ambiguity, which is  concealed in the  evolution of the free field operators, reappears again. Apparently, the naive application of the $i\e$ prescription does not solve the problem consistently. 

\section{Path Integral Approach}

Our discussion in the previous section shows that the field quantization on cosmological backgrounds requires a more rigorous analysis of the temporal asymptotics. In this section we study the problem by using path integrals and for completeness we first give a brief derivation of the in-in path integral method following \cite{j}. 

The transition amplitude in the presence of an external source $J$ from an initial field space eigenstate $\left.|\f_0\right>$ at time $t_0$ to a final field space eigenstate $\left.|\f_*\right>$ at a later time $t_*$ is given by the following standard path integral
\be\label{io}
\left<\f_*|\f_0\right>_J=\int {\cal D}\f\,\exp\left(iS[\f]+\int d^4 x\, J\,\f\right)
\ee
where the sum is over all field configurations satisfying $\f(t_0)=\f_0$ and $\f(t_*)=\f_*$. By differentiating this generating functional with respect to $J$ one can get the Green functions of time ordered field variables inserted in between the states  $\left.|\f_0\right>$ and $\left.|\f_*\right>$. 

For in-in expectation values one needs the identity operator expressed in the field configuration space at a fixed time, say $t_*$, as
\be
I=\int D\f_* \left|\f_*\right>\left<\f_*\right|.
\ee
where the integral is over all field variables at constant $t_*$. Here, one should pay attention to two different path integral measures ${\cal D\f}$ and $D\f$; while the first  corresponds to the sum over all field configurations depending on both space and time, the second represents the integration of all spatial field variables at a fixed time. For any given state $\left.|\psi\right>$ at time $t_0$, the corresponding wave-functional $\psi[\f_0]=\left<\f_0\right|\left.\psi\right>$ can be introduced by the identity 
\be\label{pc}
\left.|\psi\right>=\int D\f_0 \,\psi[\f_0]\, \left|\f_0\right>.
\ee
Using \eq{io} and its conjugate, one may now define the in-in generating functional 
\bea
Z[J^+,J^-]&&=\int D\f_*\left<\psi|\f_*\right>_{J^-}\left<\f_*|\psi\right>_{J^+}\label{z}\\
&&=\int D\f_*{\cal D\f^+}{\cal D\f^-}\exp\left(iS[\f^+]-iS[\f^-]+\int d^4 x\, [J^+\,\f^+ + J^-\f^-]\right)\psi[\f_0^+]\psi^*[\f^-_0],\nn
\eea
where $\int{\cal D\f^\pm}$ stands for the integral over all $\f^\pm$ fields defined in the interval $(t_*,t_0)$ satisfying $\f^+(t_*)=\f^-(t_*)$. It is easy to find the operator equivalent of the in-in path integral by observing that $\f^-$ and $\f^+$ insertions must be anti-time and time ordered, respectively, and $\f^-$ insertions must be placed to the left of $\f^+$ ones. For instance, the in-in path integral of $\f^+(t_1)\f^+(t_2)\f^-(t_3)\f^-(t_4)\f^-(t_5)$ gives 
\be 
\left<\psi\right|\overline{T}\left[\f(t_3)\f(t_4)\f(t_5)\right]T\left[\f(t_1)\f(t_2)\right]\left|\psi\right>,
\ee
where $\overline{T}$ denotes anti-time ordering. The normalization of the generating functional $Z[0,0]=1$ is guaranteed by the definition \eq{z}. As noted in \cite{ak1} it is possible to carry out $D\f_*$ integral explicitly, which just produces the additional condition $\dot{\f}^+(t_*)=\dot{\f}^-(t_*)$ for the ${\cal D\f^\pm}$ path integrals in \eq{z}. The proper treatment of the $D\f_*$ integral is crucial in obtaining the correct in-in propagators \cite{wein1} (see also \cite{ak2}). Below, we will see that the vacuum wave-functionals in \eq{z} also play a significant role. 

\subsection{Free Theory}

Let us now carry out the in-in path integral for the free scalar. The vacuum wave-functional at $t_0$ is given by \eq{v1}. For finite $t_0$, \eq{v1} is different than the state annihilated by $\ak$, which can be found as   
\be\label{v2}
\tilde{\Psi}_0[\f,t_0]=\tilde{C}\exp\left[\fr{i}{2} \int d^3k\,a(t_0)^3\,\fr{\dot{\f}_k^{*}(t_0)}{\f_k^{*}(t_0)}\,\f(\vc)\,\f(-\vc)\right].
\ee
To see $\ak\tilde{\Psi}_0[\f,t_0]=0$ and $\ak\Psi_0[\f,t_0]\not=0$, one can solve $\ak$ from \eq{fa} as 
\be\label{ak}
\ak=-i\int\fr{d^3 x}{(2\pi)^{3/2}}\,e^{-i\vc.\vx}\left[\dot{\f}_k^*(t_0)a(t_0)^3\f(\vx,t_0)-\f_k^{*}(t_0)P_\f(\vx,t_0)
\right]
\ee
and use the function space representation of the momentum operator $P_\f(\vx,t_0)=-i\d/\d\f(\vx,t_0)$. We should emphasize that both \eq{v1} and \eq{v2} are instantaneous states defined at time $t_0$ and one should solve the Schr\"{o}dinger equation to determine their time evolution. On the other hand, the Bunch-Davies mode function by definition obeys 
\be
\lim_{t_0\to-\infty}\fr{\dot{\f}_k^{BD}{}^*(t_0)}{\f_k^{BD*}(t_0)}\to i\fr{k}{a(t_0)},
\ee 
thus both $\Psi_0$ and $\tilde{\Psi}_0$ asymptote, up to normalization, to the same wave-functional   
\be\label{av}
\exp\left[-\fr{1}{2} \int d^3k\,a(t_0)^2k\,\f(\vc)\,\f(-\vc)\right]
\ee
in the $t_0\to-\infty$ limit.

To make the asymptotic form of the wave-functional \eq{av} manageable in the path integral \eq{z}, one may utilize the following identity  \cite{wein2}, valid for any function $f$ and time $t_*$, 
\be\label{ew}
\lim_{\e\to0}\,\int_{-\infty}^{t_*}\,\e\,e^{\e\, t}\,f(t)\,dt=f(-\infty),
\ee
which can easily be verified by partial integration. This can be used to express the instantaneous value of the function at $t_0=-\infty$ in the exponential of \eq{av} as a time integral having the same range with the other terms in the action. Consequently, the quadratic in-in path integral \eq{z} for the free scalar which is released in its Bunch-Davies vacuum at $t_0=-\infty$ can be written\footnote{As shown in \cite{ak1}, it is legitimate to apply timelike integration by parts in the action while evaluating the in-in path integral \eq{z}, which is not trivial because of the finite boundary at $t_*$.} as (for ease of notation we mostly suppress the space part of the four dimensional space-time integrals $\int d^3x$, which can easily be reinstated)  
\be\label{zz}
Z[J^+,J^-]= \lim_{\e\to0}\,\int {\cal D} \Phi \exp\left[ \int_{-\infty}^{t_*} dt\, \left( -\fr{i}{2}\,a(t)^3\, \Phi^T {\bf L}^\e\Phi+\Phi^T{\bf J}\right)\right],
\ee
where 
\be\label{j}
{\bf L}^\e =\left[\begin{array}{cc}L^\e_+&0\\0&-L^\e_-\end{array}\right],\hs{5}\Phi=\left[\begin{array}{c}\phi^+ \\\phi^- \end{array}\right], \hs{5} {\bf J}=\left[\begin{array}{c}J^+ \\ J^- \end{array}\right],
\ee
${\cal D}\Phi$ collectively denotes the in-in path integral measure and the operators $L^\e_\pm$ can be found in the momentum space as
\be\label{L}
L_\pm^\e=\fr{d^2}{dt^2}+3H\fr{d}{dt}+m^2+\fr{k^2}{a(t)^2}\mp i\,\e\, \fr{k}{a(t)}. 
\ee
The last $\e$ dependent term arises from the vacuum wave-functionals worked out by \eq{ew} where in de Sitter space the exponential $e^{\e\, t}$ can be absorbed in the scale factor $a(t)=a_0e^{H t}$. 

Since one takes $\e\to0$ limit at the end, the presence of $i\e$ terms can only affect the asymptotic behavior. This is an important property which allows one to choose different functions multiplying $\e$ terms as long as they give the same asymptotic structure. 

The Gaussian in-in path integral \eq{zz} can be calculated by applying the standard rules. Defining the inverse of ${\bf L}^\e$ in \eq{j} as
\be\label{d}
\Delta_\e=\left[\begin{array}{cc}\Delta^{++}_\e&\Delta^{+-}_\e\\ \Delta^{-+}_\e&\Delta^{--}_\e\end{array}\right],
\ee
which obeys
\be\label{gd}
{\bf L}^\e\Delta_\e=\fr{1}{a^3}\left[\begin{array}{cc}\d^4(x-x')&0\\0&\d^4(x-x')\end{array}\right],
\ee
\eq{zz} can be evaluated as 
\be\label{fr}
Z[J^+,J^-]= \lim_{\e\to0}\,\exp\left[-\fr{i}{2}\int_{-\infty}^{t_*}\int_{-\infty}^{t_*}dt\, dt'\, {\bf J}^T(t)\Delta_\e(t,t') {\bf J}(t')\right],
\ee
where the subtle $D\f_*$ integral additionally imposes \cite{wein1}  
\bea
&&\Delta_\e^{++}(t_*,t)=\Delta_\e^{-+}(t_*,t),\nn\\
&&\Delta_\e^{--}(t_*,t)=\Delta_\e^{+-}(t_*,t).\label{bct*}
\eea
Note that without loss of any generality one may take a symmetric kernel obeying $\Delta_\e^{-+}(t,t')=\Delta_\e^{+-}(t',t)$. 

It is possible to determine $\Delta_\e$ in terms of the mode functions of \eq{L}. In de Sitter space they can be solved exactly in terms of the Whittaker functions, but for our purposes it is enough to fix their early time asymptotic forms. To the leading order as $a\to0$, while the two solutions of $L^\e_+\m=0$ can be fixed like
\bea
&&\m_1=\fr{1}{a\sqrt{2k}}\exp\left(\fr{ik}{aH}\right)\left[\fr{k}{aH}\right]^{\e/(2H)}\left\{1+{\cal O}(a)\right\},\nn\\
&&\m_2=\fr{1}{a\sqrt{2k}}\exp\left(\fr{-ik}{aH}\right)\left[\fr{k}{aH}\right]^{-\e/(2H)}\left\{1+{\cal O}(a)\right\},\label{lm}
\eea
the two solutions of $L_-^\e\n=0$ can be found as   
\bea
&&\n_1=\fr{1}{a\sqrt{2k}}\exp\left(\fr{ik}{aH}\right)\left[\fr{k}{aH}\right]^{-\e/(2H)}\left\{1+{\cal O}(a)\right\},\nn\\
&&\n_2=\fr{1}{a\sqrt{2k}}\exp\left(\fr{-ik}{aH}\right)\left[\fr{k}{aH}\right]^{\e/(2H)}\left\{1+{\cal O}(a)\right\}. \label{ln}
\eea
Like the usual mode functions, we normalize these to obey the standard Wronskian relation $\m_1\dot{\m}_2-\dot{\m}_1\m_2=\n_1\dot{\n}_2-\dot{\n}_1\n_2=i/a^3$ (the Wronskian can be found from \eq{L} to be proportional to $1/a^3$). Furthermore, they are also chosen to satisfy 
\bea
&&\lim_{\e\to0}\m_1=\lim_{\e\to0}\n_1=\f^{BD}_k,\nn\\
&&\lim_{\e\to0}\m_2=\lim_{\e\to0}\n_2=\f_k^{BD}{}^*,\
\eea
where $\f^{BD}_k$ is the de Sitter Bunch-Davies mode function \eq{bdm}. Introducing two {\it auxiliary} scalar fields $\f^\pm$ as in \eq{fa} where the mode functions are chosen to be $(\m_1,\m_2)$ for $\f^+$ and  $(\n_1,\n_2)$ for $\f^-$, the in-in propagator $\D_\e$ can be constructed as 
\bea
&&\Delta_\e^{++}(t,\vx,t',\vx\,')=i\vvc \right. T\f^+(t,\vx)\f^+(t',\vx\,')\left. \vv,\nn\\
&&\Delta_\e^{--}(t,\vx,t',\vx\,')=i\vvc \right. \overline{T}\f^-(t,\vx)\f^-(t',\vx\,')\left.\vv,\label{cf}\\
&&\Delta_\e^{-+}(t,\vx,t',\vx\,')=i\vvc \right. \f^-(t,\vx)\f^+(t',\vx\,')\left.\vv,\nn\\
&&\Delta_\e^{+-}(t,\vx,t',\vx\,')=i\vvc \right. \f^-(t',\vx\,')\f^+(t,\vx)\left.\vv.\nn
\eea
It can now be checked directly that \eq{cf} obeys \eq{gd}. Strictly speaking, the boundary conditions \eq{bct*} are violated because of the asymmetry of the mode functions. However, this problem can either be solved by giving a tiny time dependence to $\e$ so that $\e(t_*)=0$ or by modifying the mode functions \eq{lm} and \eq{ln} by the terms that will disappear as $\e\to0$.  

Note that for $\e\not=0$ the above functions {\it cannot} be interpreted as the expectation values of the scalar field \eq{fa}. Such an interpretation requires $\Delta_\e^{++}(t,t)=\Delta_\e^{--}(t,t)=\Delta_\e^{-+}(t,t)$ since each supposed to yield the expectation value of the same operator, i.e. two $\f$ fields at the coincident time. One may check that it is impossible to satisfy this condition with the mode functions \eq{lm} and \eq{ln}, or with their linear combinations. When $\e\not=0$, the $\pm$ branches corresponding to forward and backward time evolution become asymmetric which suggests that some form of {\it non-unitarity} is introduced.  

For any fixed finite time, the auxiliary fields $\f^\pm$ become equal to the original scalar field $\f$ for vanishing $\e$, i.e.   
\be
\lim_{\e\to0}\f^{\pm}(t,\vx)=\f(t,\vx),\hs{5}\textrm{finite}\hs{2}t.
\ee
Then, all correlation functions \eq{cf} yield the standard results. However, there is a crucial difference between taking the $\e\to0$ limit {\it before or after} sending a time argument to $-\infty$. To underline this difference, one can use  \eq{lm} and \eq{ln} in \eq{cf} by paying attention to the time orderings to see that in the limit $\e\to0$ following relations hold
\bea
&&\Delta_\e^{++}(t,t')= e^{-\e|t-t'|/2}\,\Delta^{++}_{\e=0}(t,t'),\hs{5}\Delta_\e^{-+}(t,t')= e^{\e(t+t')/2}\,\Delta^{-+}_{\e=0}(t,t'),\nn\\
&&\Delta_\e^{--}(t,t')=e^{-\e|t-t'|/2}\,\Delta_{\e=0}^{--}(t,t'),\hs{5}\Delta_\e^{+-}(t,t')= e^{\e(t+t')/2}\,\Delta^{+-}_{\e=0}(t,t').\label{et}
\eea
Clearly, letting $\e\to0$ before sending $t\to-\infty$ or $t'\to-\infty$ gives the canonical correlators, which are plagued by the infinite phase issue. In that case, the path integral approach does not offer anything new.  

Yet, one may prefer to keep $\e$ finite while letting any of the time arguments to $-\infty$, which then implies by \eq{et} that the Green functions pick up exponentially decreasing factors. This behavior is precisely the one sought for the convergence of the time integrals in the in-in perturbation theory. As noted above, keeping $\e\not=0$ introduces some form of non-unitarity  and thus in this second option the temporal past infinity limit of the Green functions is actually carried out under this condition. 

To understand the nature of this non-unitary evolution, consider the following initial state at time $t_0$
\be\label{02}
\left.|\psi\right>=c\left.\vv+\int d^3 k\, g(\vc)\,\akd\left.\vv,\hs{5}\left<\psi|\psi\right>=1,
\ee
where $\left.\vv$ is the vacuum defined by $\ak\left\vv=0$. Using \eq{v2} and the conjugate of \eq{ak} giving $\ak^\dagger$, the corresponding wave-functional can be found as 
\be\label{ex}
\psi[\f_0,t_0]=\left[c+\int d^3 k\, \fr{g(\vc)}{\f_k^*(t_0)}\,\f_0(-\vc)\right]\tilde{\Psi}_0[\f_0,t_0],
\ee
where the argument of the functionals is the spatial field variable $\f_0(\vx)$ at $t_0$ and the Fourier transformed variable $\f_0(\vc)$ is defined as in \eq{ft} (in \eq{ex} the distinction between the mode function $\f_k(t_0)$ and the field variable $\f_0(-\vc)$ must be evident). Let us now work out the following path integral 
\be
\int D\f_*{\cal D\f^+}{\cal D\f^-}e^{iS[\f^+]-iS[\f^-]}\,\f^+(t,\vec{k}_1)\f^+(t,\vec{k}_2)\,\psi[\f^+_0,t_0]\psi^*[\f^-_0,t_0],\label{exp}
\ee
which gives the two point function $\left<\psi|\right.\hs{-1}\f(t,\vec{k}_1)\f(t,\vec{k}_2)\hs{-1}\left.|\psi\right>$. 
Using \eq{ex} it can be expressed as the in-in Gaussian path integral of
\be\label{gi}
\f^+(t,\vec{k}_1)\f^+(t,\vec{k}_2)\,\left[c+\int d^3 k\, \fr{g(\vc)}{\f_k^*(t_0)}\,\f^+(t_0,-\vc)\right]\left[c^*+\int d^3 \tilde{k}\, \fr{g^*(\tilde{k})}{\f_{\tilde{k}}(t_0)}\,\f^{-}(t_0,-\tilde{k})^*\right],
\ee
with the weight function $\tilde{\Psi}_0[\f_0,t_0]$ given in \eq{v2}. As before, one can use the identity \eq{ew} to handle the vacuum wave-functional as $t_0\to-\infty$ giving the $\e$ dependent operators \eq{L}. Therefore, \eq{gi} can be evaluated by using  Wick's theorem and the corresponding two point functions \eq{cf}, where one chooses $\f_k=\f_k^{BD}$. 

If one naively sets  $\e=0$ before taking the $t_0\to-\infty$ limit, Wick contractions give
\be\label{sta} 
|\f_{k_1}(t)|^2\d^3(\vec{k}_1+\vec{k}_2)+g^*(-\vec{k}_2)g(\vec{k}_1)\f_{k_1}(t)\f_{k_2}^*(t)+g^*(-\vec{k}_1)g(\vec{k}_2)\f_{k_2}(t)\f_{k_1}^*(t),
\ee
where we have used $\left<\psi|\psi\right>=1$ that imposes $|c|^2+\int d^3k |g|^2=1$. Here, one may see that the mode functions in the denominators in \eq{gi} are {\it exactly} canceled out by the terms arising from the contractions of $\f^+(t_0,-\vc)$ and  $\f^{-}(t_0,-\tilde{k})^*$. Moreover, \eq{sta} is the standard result that would follow from using \eq{fa} directly in $\left<\psi|\right.\f(t,\vec{k}_1)\f(t,\vec{k}_2)\left.|\psi\right>$. 

On the other hand, if one insists keeping $\e$ nonzero till the end of taking the $t_0\to-\infty$ limit, then all contractions of $\f^+(t_0,-\vc)$ and  $\f^{-}(t_0,-\tilde{k})^*$ in \eq{gi}, {\it including their own}, vanish by the  decaying $\e$ dependent terms in \eq{et} giving 
\be
|c|^2\,|\f_{k_1}(t)|^2\d^3(\vec{k}_1+\vec{k}_2).
\ee
Therefore, this alternative approach exactly corresponds to projecting the initial state \eq{02} onto the ground state $\left.\vv$. Note, however, that the final result depends on $|c|^2$, i.e. the overlap of $\left.|\psi\right>$ and $\left.\vv$. 

It is not difficult to see that this procedure generally projects onto the ground state $\left.\vv$ in the expectation values for any given initial state in the Fock space spanned by the vectors $\left.\vv$, $\ak^\dagger\left.\vv$, $a_{\vec{k}_1}^\dagger a_{\vec{k}_2}^\dagger\left.\vv$ ... that are properly normalized by test functions. The crucial point here is that the contraction of any field variable having finite time argument with any other field variable defined at past infinity coming from the initial state wave-functional vanishes when $\e\not=0$ because of the $\e$ dependent decaying factors in \eq{et}. Although $(++)$ and $(--)$ inner contractions of state wave-functional field variables do not vanish since $\Delta_\e^{++}(t,t')$ and $\Delta_\e^{--}(t,t')$ in \eq{et} do not decay as $t,t'\to-\infty$ while $|t-t'|$ is fixed, this only affects the overall normalization of the correlation function. 

Lastly, we would like to discuss an alternative way of obtaining the above projection, which will be useful for our subsequent considerations. Assume that the states in our problem are evolved by the operator $(1-i\e)H_0$ rather than the Hamiltonian $H_0$ itself. Note that this is not the same as deforming the time integration contour; the explicit time dependence of $H_0$ is kept intact here and the Hamiltonian is simply scaled by a complex number. It is easy to utilize this new time evolution in the phase space in-in path integral by the substitution 
\be\label{hr}
H_0\to(1-i\e)H_0
\ee
in the action. Let us for the moment ignore the presence of the wave-functionals and carry out the Gaussian momentum integral, which gives the following action for $\pm$ branches 
\be
S^{\pm}=\int\,d^4 x\, a^3\,(1\pm i\e) \left[\fr12\dot{\f}^\pm{}^2-(1\mp i\e)^2\left(\fr{1}{2a^2}\del_i\f^\pm\del_i\f^\pm+\fr12 m^2\f^\pm{}^2\right)\right],
\ee
where we have used $1/(1\pm\e)\to(1\mp\e)$ because only the leading order $\e$ terms matter. Since $\e$ contributions are only important for the asymptotic behavior, one may keep the $\e$ dependence of the most singular term as $a\to0$, which is $(\del\f)^2/a^2$. Consequently, the following operators show up in the path integral  
\be\label{L2}
L_\pm^\e=\fr{d^2}{dt^2}+3H\fr{d}{dt}+m^2+(1\mp i\e)^2\fr{k^2}{a(t)^2},
\ee
which should be compared to \eq{L}. The modes of \eq{L2} are identical to the Bunch-Davies mode functions where one replaces $k\to k(1\mp i\e)$ for $\pm$ branches, respectively. It is then straightforward to see that the Fourier transformed components\footnote{Note that all the kernels $\Delta^{\pm\pm}_\e(t,\vx,t',\vx\,')$ are actually functions of the difference $(\vx-\vx\,')$ so their Fourier transformation can be defined as $\Delta^{\pm\pm}_\e(t,\vx,t',\vx\,')=(2\pi)^{-3/2}\int d^3k\,e^{i\vc.(\vx-\vx')} \Delta_\e^{\pm\pm}(\vc,t,t')$.\label{foot6}} of the Bunch-Davies correlators obey
\bea
&&\Delta_\e^{++}(\vc,t,t')= e^{-\e k|\eta-\eta'|}\,\Delta^{++}_{\e=0}(\vc,t,t'),\hs{5}\Delta_\e^{-+}(\vc,t,t')= e^{\e k (\eta+\eta')}\,\Delta^{-+}_{\e=0}(\vc,t,t'),\nn\\
&&\Delta_\e^{--}(\vc,t,t')=e^{-\e k|\eta-\eta'|}\,\Delta_{\e=0}^{--}(\vc,t,t'),\hs{5}\Delta_\e^{+-}(\vc,t,t')= e^{\e k(\eta+\eta')}\,\Delta^{+-}_{\e=0}(\vc,t,t'),\label{et2}
\eea
where the conformal time in de Sitter space is defined as usual by $a(t)=-1/(H\eta)$. We thus see that the replacement \eq{hr} also gives the desired falloff of the Green functions like \eq{et}. If now an arbitrary initial wave-functional at past infinity is included in the path integral, only the vacuum piece would yield a nonzero contribution and other components are projected out (if $\e$ is kept nonzero till the end of the calculation), which can be verified as before by expanding the state in the Fock space basis and then using \eq{et2} together with Wick's theorem.

Summarizing, we have shown that it is possible to make sense of past eternal time evolution in the path integral approach either by carefully treating the $i\e$ terms coming from the vacuum wave-functional or by deforming the Hamiltonian of the system by an extra $i\e$ piece as in \eq{hr}. In both cases, insisting on keeping $\e$ nonzero introduces non-unitary evolution, which eventually projects an arbitrary initial state onto the vacuum at past infinity. 

\subsection{Interacting Theory} \label{3b}

It is not difficult to deal with the interactions perturbatively by using the above results. Including the interaction potential $V(\f)$ in the action, the in-in path integral can be calculated order by order in powers of $V(\f)$ by expanding the exponential. As before, we take an operator that has a single time argument $O(t_*)$, which can be included as a $+$ branch variable $O^+(t_*)$ in the path integral (while studying more general cases is also straightforward). For each term in the perturbative expansion, the corresponding operator equivalent can be found by paying attention to time orderings of $\pm$ variables and it is not hard to see that the series is equivalent to \eq{inp} where $H_I=a^3\int d^3 x\, V(\f)$ (of course the vacuum expectation values must be calculated as in the previous subsection which is the new ingredient). For example, the first order correction becomes the path integral of the following term
\be\label{f4}
i\int_{-\infty}^{t_*}\,d^3 x\, dt\, a(t)^3\,\left(O^+(t_*)V(\f^-(t,\vx))-O^+(t_*)V(\f^+(t,\vx))\right)
\ee
which has the operator counterpart $i\int_{-\infty}^{t_*}d^3 x\, dt\, a(t)^3\,[V(\f(t,\vx)),O(t_*)]$ that equals the $N=1$   term in \eq{inp}. Note that the time integration variable is not complexified here thus the previously mentioned problem about the asymmetry of the contours ruining the commutator structure does not arise. After expanding up to the desired order, one can apply Wick's theorem to evaluate the path integral. 

\begin{figure}
\centerline{\includegraphics[width=9cm]{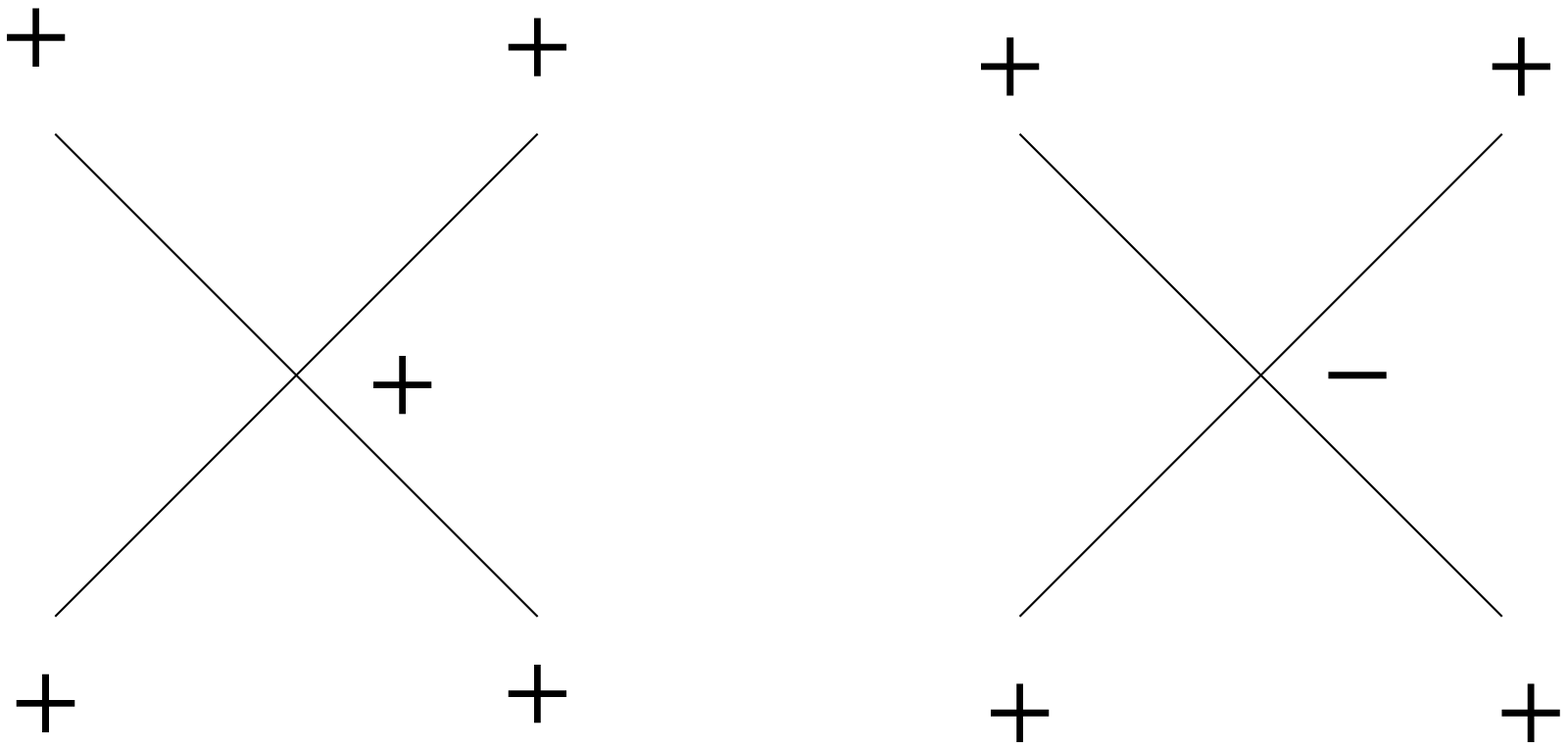}}
\caption{The in-in graphs corresponding to the tree level correction to $\left<\right.\hs{-1}\f(t_*,\vc_1)...\f(t_*,\vc_4)\hs{-1}\left.\right>$ in $\l\f^4$ theory. There are two vertices corresponding to $\l(\f^+)^4$ and $\l(\f^-)^4$ coming from $S^+$ and $S^-$, respectively. The operator of interest is taken as a $+$ branch variable.} 
\label{fig1}
\end{figure}

In the following we consider $V=\l\f^4$ theory to work out an explicit example  and take $O(t_*)=\f(t_*,\vc_1)\f(t_*,\vc_2)\f(t_*,\vc_3)\f(t_*,\vc_4)$. For the moment assume that the system is released in its {\it free ground state} that gives the propagators \eq{cf}. The first order correction to $\left<O(t_*)\right>$ can be found by path integrating \eq{f4} using Wick's theorem that yields up to a numerical proportionality constant the following result
\bea
&&\,\l\,\d^3(\vc_1+\vc_2+\vc_3+\vc_4)\int_{-\infty}^{t_*} dt\, a(t)^3\label{i1}\\
&&\left[\D^{++}_\e(\vc_1,t_*,t)...\D^{++}_\e(\vc_4,t_*,t)-\D^{-+}_\e(\vc_1,t,t_*)...\D^{-+}_\e(\vc_4,t,t_*)\nn
\right],
\eea
which can be pictured as in Fig. \ref{fig1}. We are concerned about the convergence of this time integral near $t=-\infty$. Indeed, from \eq{lm}, \eq{ln} and \eq{cf} one may see that 
\be
\lim_{t\to-\infty}\D^{++}_\e(\vc,t_*,t)\propto\fr{1}{a(t)}e^{-ik/Ha(t)}e^{\e t/2},\hs{5}\lim_{t\to-\infty}\D^{-+}_\e(\vc,t_*,t)\propto\fr{1}{a(t)}e^{ik/Ha(t)}e^{\e t/2},
\ee
therefore the integral has the following form in the conformal time near infinity 
\be\label{74}
\int_{-\infty}d\eta\, e^{iK\,\eta}\,(-\eta)^{-\e},
\ee
where $K=\pm(k_1+k_2+k_3+k_4)$ corresponds to the first and the second terms in \eq{i1}. Eq. \eq{74} becomes convergent only when $\e\not=0$ and moreover one can safely set $\e$ to zero after carrying out the integral like in the calculus of distributions (indeed we are dealing with a distribution in the momentum space). If instead one uses \eq{et2} as the asymptotic form of the in-in propagators, the integral \eq{i1} near infinity becomes 
\be\label{75}
\int_{-\infty}d\eta\, e^{iK\,\eta+\e |K|\,\eta},
\ee
which is equivalent to assigning an appropriate imaginary piece to the conformal time. Either with \eq{74} or \eq{75} one gets the same result after setting $\e=0$, which is an important consistency requirement. 

Let us now assume that the system is released, {\it not} in its free ground state, but in a general state like \eq{02} in the Fock space. In that case, in addition to \eq{f4} one has extra terms in the path integral coming from the state wave-functional as the ones in the square brackets in \eq{gi}; all are integrated out in the presence of the Gaussian vacuum wave-functional. Applying again Wick's theorem one sees that any contraction of a term in $O^+$ or in $V(\f^\pm)$ with $\f^{\pm}(-\infty,\vc)$ coming from the initial state vanishes by the asymptotics of the two point functions.\footnote{For the validity of this statement for $(++)$ and $(--)$ contractions involving the potential $V(\f^\pm)$ and the field variable coming from the initial state $\f^{\pm}(-\infty,\vc)$, the structure of the corresponding Green functions in \eq{et} shows that one must first carry out the limit while treating the other time argument as a finite quantity, despite it later becomes an integration variable extending to infinity.} Therefore, the $i\e$ prescription also projects onto the free ground state in the presence of interactions. 

One may be bothered by the fact that the $i\e$ prescription still projects onto the free ground state $\left.\vv$ rather than the full interacting one $\left.|\O\right>$. This seems inevitable in perturbation theory as we will discuss in the next section. Still, by comparing the terms in the Hamiltonian \eq{ham} one may conclude that (in the presence of an IR cutoff) the gradient term eventually dominates the potential $V(\f)$ as $a\to0$ 
which then implies     
\be
\lim_{a\to0} \left.|\O\right>\to\left.\vv.
\ee
Thus, for the self-interacting scalar field in de Sitter space the free and exact {\it instantaneous} vacua must coincide at {\it past asymptotic infinity,} at least in perturbation theory where $\left.\vv$ and $\left.|\O\right>$  do not differ too much anyway. This property may clearly fail when derivative interactions present and in that case the $i\e$ prescription may not provide a projection onto the interacting vacuum state. 

\section{Operator Formalism}

In this section, our aim is to apply the $i\e$ prescription in the in-in theory using the operator formalism. As discussed in subsection \ref{2b}, this is normally implemented by complexifying the time variable of the interaction picture unitary evolution operator as in \eq{ue2}. However, this procedure is problematic and not well justified as pointed out before. Here, we study a closely related and technically more appropriate approach suggested by the path integral considerations above. Namely, we consider the evolution of the states dictated by $(1-i\e)H$, where $H$ is the {\it full Hamiltonian}, which gives the following modified Schr\"{o}dinger equation 
\be\label{iH}
i\fr{\del\psi}{\del t}=(1-i\e)H\,\psi. 
\ee
Note that this deformation, which is simply multiplying $H$ by a complex number, is completely different than complexifying the time variable. Note also that we apply the scaling to the full $H$, not just to the interaction Hamiltonian $H_I$. As we will see, \eq{iH} gives a well defined time evolution from past asymptotic infinity and further involves a projection. 

\subsection{A Sample Calculation in the Free Theory}\label{sample}

To warm up we first examine the solutions of \eq{iH} in the free theory that has the form of a Gaussian wave-functional (our analysis is similar to \cite{s1,s2} which study squeezing of states in inflation). In the function space representation, the free Hamiltonian becomes 
\be
H_0(t)=\fr12 \int d^3x \left[-\fr{1}{a(t)^3}\fr{\d^2}{\d\f(\vx)^2}+\,a(t)\, (\del_i\f(\vx))^2+a(t)^3m^2\f(\vx)^2\right].\label{ham0}
\ee
One can check that the following state 
\be\label{v22}
\tilde{\Psi}[\f,t]=\tilde{C}e^{-i\int_{t_0}^tE_0(t')dt'}\exp\left[\fr{i}{2} \int d^3k\,a(t)^3\,\fr{\dot{\f}_k^*(t)}{\f_k^*(t)}\,\f(\vc)\,\f(-\vc)\right]
\ee
solves the Schr\"{o}dinger equation 
\be
i\fr{\del\tilde{\Psi}}{\del t}=H_0(t)\tilde{\Psi},
\ee
where $\f_k$ obeys the standard mode equation \eq{me}, 
\be
E_0(t)=-\fr12\fr{{\cal V}}{(2\pi)^3}\int d^3k\,\fr{\dot{\f}_k^*(t)}{\f_k^*(t)}
\ee
and ${\cal V}=\int d^3x$ is the volume of the space. Note that $\tilde{\Psi}[\f,t]$ coincides with \eq{v2} at $t=t_0$  and also that $\ak\tilde{\Psi}[\f,t]\not=0$ unless $t=t_0$, where $\ak$ is given by \eq{ak}. Therefore, in the limit $t_0\to-\infty$, the state \eq{v22} with $\f_k=\f_k^{BD}$ can be viewed as the time evolved vacuum of past infinity.  

Consider now the following general Gaussian state 
\be\label{pe}
\Psi_\e[\f,t]=Ce^{-i\th(t)}\exp\left[-\fr{1}{2} \int d^3k\,\s(k,t)\,\f(\vc)\,\f(-\vc)\right],
\ee
which evolves according to 
\be\label{iHe}
i\fr{\del\Psi_\e}{\del t}=(1-i\e)H_0(t)\Psi_\e. 
\ee
It is an easy exercise to see that \eq{iHe} implies
\bea
&&\dot{\th}=(1-i\e)E(t),\hs{5}E(t)=\fr{{\cal V}}{2(2\pi a)^3}\,\int d^3 k\, \s(k,t),\nn\\
&&i\dot{\s}(k,t)+a^3(1-i\e)\left[m^2+\fr{k^2}{a^2}-\fr{\s(k,t)^2}{a^6}\right]=0.\label{sth}
\eea
Guided by the form of \eq{v22}, one may define 
\be\label{sl}
\s(k,t)=-i\fr{a^3}{(1-i\e)}\,\fr{\dot{\l}_k}{\l_k}
\ee
to see that \eq{sth} is satisfied if
\be\label{le}
\ddot{\l}_k+3H\dot{\l}_k+(1-i\e)^2\left(m^2+\fr{k^2}{a^2}\right)\l_k=0.
\ee
This relates the solution of the first order nonlinear non-homogeneous differential equation \eq{sth} to a second order linear homogeneous one \eq{le}. Although there appear two integration constants from \eq{le}, only one combination of them matters for $\s(k,t)$ since it just involves the ratio of $\dot{\l}_k$ and $\l_k$.  

Assume that the Gaussian state is {\it fixed} at $t_0$ by a given variance $\b(k)$, i.e. $\s(k,t_0)=\b(k)$. Without loss of any generality one may set $\l_k(t_0)=1$ and determine $\dot{\l}_k(t_0)$ from \eq{sl} so that 
\be\label{lic}
\l_k(t_0)=1,\hs{5} \dot{\l}_k(t_0)=i(1-i\e)\b(k)/a(t_0)^3.
\ee
Eq. \eq{le} can now be solved uniquely with these initial conditions, which then determines by \eq{sl} the time evolution of the state under \eq{iHe}. In de Sitter space, the two solutions of \eq{le} can be expressed in terms of the Bunch-Davies mode functions $\f^{BD}_k$ and $\f_k^{BD}{}^*$ given in \eq{bdm} by replacing $m\to(1-i\e)m$ and $k\to(1-i\e)k$. Let us denote those by $f_1$ and $f_2$, respectively, and fix their Wronskian for convenience to be $f_1\dot{f}_2-\dot{f}_1f_2=1/(2k(i+\e)a^3)$. Writing the general solution of \eq{le} as
\be\label{lcc}
\l_k(t)=c_1(t_0)f_1(t)+c_2(t_0)f_2(t).
\ee
the constants $c_1$ and $c_2$ satisfying \eq{lic} can be solved as
\bea
&&c_1(t_0)=a(t_0)^3\dot{f}_2(t_0)-i(1-i\e)\b(k)f_2(t_0),\nn\\
&&c_2(t_0)=-a(t_0)^3\dot{f}_1(t_0)+i(1-i\e)\b(k)f_1(t_0).\label{cc}
\eea
Using $\l_k$ in \eq{sl} gives the general solution $\s(k,t)$ with the initial condition $\s(k,t_0)=\b(k)$. 

Let us now consider the limit $t_0\to-\infty$, i.e. assume that the Gaussian state is released from past infinity. One can see by using the large time asymptotics of the Hankel functions that $f_1(t_0)\to \exp[\e k/a(t_0)H]$ and  $f_2(t_0)\to \exp[-\e k/(a(t_0)H)]$ as $t_0\to-\infty$ owing to the fact that $f_1$ and $f_2$ are obtained from $\f^{BD}_k$ and $\f^{BD}_k{}^*$ in \eq{bdm} by the scaling $k\to(1-i\e)k$. Consequently \eq{cc} implies $c_1(t_0)\to0$ and $c_2(t_0)\to\infty$ as $t_0\to-\infty$. Keeping $\e$ nonzero is crucial in taking this limit, otherwise both $f_1(t_0)$ and $f_2(t_0)$ become infinitely oscillating functions rather than becoming exponentially growing and decaying ones. Using \eq{lcc} in \eq{sl}, employing the behavior of $c_1$ and $c_2$ in the limit and finally setting  $\e=0$ gives 
\be
\lim_{t_0\to-\infty}\s(k,t)=-ia(t)^3\fr{\dot{\f}_k^{BD}{}^*(t)}{\f_k^{BD}{}^*(t)}.
\ee
Up to a (possibly) divergent normalization constant this coincides with \eq{v22} where $\f_k=\f_k^{BD}$, which is the vacuum at past infinity evolved to time $t$ by the usual Schr\"{o}dinger equation. The Gaussian initial state $\Psi_\e[\f,-\infty]$ given by \eq{pe} turns into a superposed state with infinite components including the vacuum when expanded in the standard basis of the Fock space, so this is a highly nontrivial result.

It is possible to work out other examples similar to the above computation. As a nice instructive case, which illustrates the characteristics of the evolution from past infinity, one may take the state
\be\label{pe2}
\Psi_\e[\f,t]=Ce^{-i\th(t)}\,\left[\int d^3k\, f(k,t)\,\f(\vc)\right]\,\exp\left[-\fr{1}{2} \int d^3k\,\s(k,t)\,\f(\vc)\,\f(-\vc)\right]
\ee
that depends on two functions $f(k,t)$ and $\s(k,t)$. Using \eq{pe2} in \eq{iHe} one obtains the same equations in \eq{sth} together with
\be
i\dot{f}(k,t)-\fr{(1-i\e)}{a(t)^3}f(k,t)\s(k,t)=0,
\ee
which can be solved as
\be\label{93}
f(k,t)=f(k,t_0)\exp\left[-i(1-i\e)\int_{t_0}^t\fr{\s(k,t')}{a(t')^3}dt'\right]. 
\ee
It is easy to see that sending $t_0\to-\infty$ while keeping both $f(k,t_0)$ and $\s(k,t_0)$ fixed yields  
\be\label{flim}
\lim_{t_0\to-\infty}f(k,t)=0.
\ee
In this case the deformed Schr\"{o}dinger equation \eq{iHe} completely kills the state evolving from past infinity.\footnote{Incidentally, using $\s(k,t)=-ia^3\dot{\f}_k^*(t)/\f_k^*(t)$ with $\e=0$ in \eq{93} gives $f(k,t)=g(k)/\f_k^*(t)$ which precisely corresponds to a one particle state, see \eq{ex}.} One should note that the prescribed state \eq{pe2} has zero overlap with the vacuum initially. 

A little care is needed for the last conclusion since one may insist keeping the norm of the state fixed while taking $t_0\to-\infty$ limit. To elucidate about this concern let us consider the superposition of the two states given in \eq{pe} and \eq{pe2}, which is still a solution. The norm of this state, which is defined by the path integral over the field configurations at a fixed time as $\left<\Psi|\Psi\right>=\int D\f\,\Psi^*[\f]\Psi[\f]$, can be found to be proportional to 
\be\label{norm}
\left[|c|^2+\int d^3 k \left|f(k,t)\right|^2\s(k,t)\right]^{1/2},
\ee
where the first and the second terms come from \eq{pe} and \eq{pe2}, respectively, and $c$ is a finite number that depends on the weight of the superposition. As a result, the normalization cannot prevent the vanishing of \eq{pe2} for such a superposed state that has some nonzero overlap with the vacuum, because \eq{norm} does not vanish while $f(k,t)\to0$. 

On the other hand, the norm of the state \eq{pe2} alone, which is given by \eq{norm} with $c=0$, also vanishes in the limit \eq{flim}. In that case, whether one gets a sensible final result by normalizing the state before taking the limit depends on the forms of the initial functions $f(k,t_0)$ and $\s(k,t_0)$. For instance if $\s(k,t_0)=\s$ that does not depend on $k$, the leading order vanishing term in \eq{93} becomes $\exp(-\e\s/3Ha(t_0)^3)$, which is also $k$ independent. This factor can be taken out of the momentum integrals in \eq{pe2} and \eq{norm}, and cancels out when the state is normalized.   

\subsection{General Case}\label{gc}

The above computation leads us to the following general conclusion: Evolving an arbitrary initial state from past infinity by \eq{iH} and setting eventually $\e=0$ yield a state  which equals, up to a normalization constant, to the vacuum of past infinity evolved up to the same time by the same method.\footnote{One of the main assertions of this work is that infinite unitary evolution produces ambiguous infinite phases. In simple cases like the one discussed below \eq{cc1}, this phase cancels out in calculating the vacuum expectation values. However, we hope the reader is convinced that the most natural and technically sound way of defining evolution from asymptotic past infinity is keeping $\e$ finite till the last moment even for the ground state.} In other words, when the arbitrary initial state is expanded in the standard Fock space basis, the evolution dictated by \eq{iH} kills all but the vacuum component at past infinity. It is easy to prove this assertion when the Hamiltonian is time independent and in the time dependent case the following heuristic argument can be given: One can introduce the complete orthonormal eigenstates of the Hamiltonian at any fixed time as
\be
H(t)\left|\psi_n(t)\right>=E_n(t)\left|\psi_n(t)\right>,\hs{5}\left<\psi_m(t)|\psi_n(t)\right>=\d_{mn}.
\ee
An arbitrary state $\left.|\psi\right>$ can always be expanded in this bases so that 
\be
\left.|\psi\right>=\sum_n c_n(t)\left.|\psi_n(t)\right>.
\ee
Using this expansion in \eq{iH} gives
\be\label{cdot}
\dot{c}_m+i(1-i\e)E_mc_m+\left<\psi_m(t)|\right.\dot{\psi}_m(t)\left.\hs{-1}\right>c_m=-\sum_{n\not=m} c_n\left<\psi_m(t)|\right.\dot{\psi}_n(t)\left.\hs{-1}\right>.
\ee
In the left hand side the number $iE_m+\left<\psi_m(t)|\right.\dot{\psi}_m(t)\left.\hs{-1}\right>$ multiplying $c_m$ is pure imaginary and this term can only modify the phase of $c_m$. On the other hand, the positive real term $\e E_m c_m$ gives rise to an exponential decrease $\exp(-\e \int^t  E_m(t')dt')$. The non-homogeneous contribution in the right hand side cannot change this behavior since each $c_n$ is constantly subject to a similar decay. Indeed, one may note that
\be
\fr{\del}{\del t} \left<\psi|\psi\right>=-2\e \left<\psi|H|\psi\right><0.
\ee
Therefore, the norm $\left<\psi|\psi\right>=\sum_n |c_n|^2$ decreases (more likely exponentially) under the evolution dictated by \eq{iHe}, which is consistent with the assertion that $c_n$ components fall off exponentially in time. Since the energy of the ground state is the lowest for all times, i.e. $E_0(t)\leq E_m(t)$, $c_0$ must be the least decreasing component.

Because we are interested in the $\e\to0$ limit eventually, the $\e$ deformation can only alter the asymptotic behavior (any finite quantity multiplied by $\e$ finally vanishes). Besides, even though the norm of the vector decreases, what really matters is the relative magnitude of the components since the norm can be always set to unity. Take now a state specified at $t_0$ with components $c_n(t_0)$, evolve it to a later time $t$ and take the limit $t_0\to-\infty$ while keeping $c_n(t_0)$ and $\e$ finite, and finally set $\e=0$. We can carry out this limit by evolving the state first from $t_0$ to $T$, which is some enormously negative but nevertheless finite time,  with $\e\not=0$ and consider the evolution there on from $T$ to $t$ with $\e=0$. Indeed $T$ can be chosen as $t_0+ 1/(\e E_{min})$, where $E_{min}$ is the greatest lower bound for the ground state energy $E_{min}<E_0(t)$. In that case the asymptotic time evolution in the interval $(t_0,T)$ exponentially kills all components relative to $c_0$. In the second interval $(T,t)$ one has the usual unitary time evolution that mixes the components while preserving the norm. Therefore, by carrying out the limit in this way one obtains  the same state up to normalization at time $t$ if one would have initially chosen $c_0(t_0)=1$, $c_n(t_0)=0$ for $n\not=0$,  which corresponds to the evolution of the asymptotic vacuum from $t_0=-\infty$ to $t$. 

This argument shows that if the non-unitary evolution operator in the Schr\"{o}dinger picture is introduced as
\be\label{uhe2}
U_H^\e(t)=T\exp\left[-i(1-i\e)\int_{-\infty}^t H(t')dt'\right],
\ee
one must have
\be\label{uevac}
\lim_{\e\to0}U_H^\e(t)\left.|\psi\right>=c(\psi)\lim_{\e\to0}U_H^\e(t)\left.\tvv,
\ee
where $\left.\tvv$ is the vacuum of $H(t)$ at past infinity, $\left.|\psi\right>$ is an arbitrary state that has some overlap with $\left.\tvv$ and $c(\psi)$ is a state dependent constant (when $\left<\psi\tvv=0$ one may naively set $c(\psi)=0$ unless the final state is normalized, which may change the limit as discussed in the explicit example at the end of subsection \ref{sample}). One should thus define, as opposed to \eq{ue}, 
\be\label{tvv2}
\tvvc\right. O\left.\tvv(t)=\lim_{\e\to0}\,\fr{\left<\psi|\right. U_H^\e(t)^\dagger \,O\,U_H^\e(t)\left.|\psi\right>}{\left<\psi|\right. U_H^\e(t)^\dagger U_H^\e(t)\left.|\psi\right>},
\ee
 where the left hand side must be understood as the  vacuum expectation value of the (Schr\"{o}dinger picture) operator $O$ at time $t$ when the system is released in its asymptotic vacuum state $\left.\tvv$ at past infinity. The denominator in \eq{tvv2} is fixed by normalization that ensures $\tvvc\right. I\left.\tvv=1$. The validity of \eq{tvv2} is evident when $H$ is time independent (a proof is actually given in \eq{po}) or when the adiabatic approximation is good (see the discussion around \eq{adi}). Moreover \eq{tvv2} has  also been validated  in the free theory using the path integral formalism in the previous section, see the paragraph of \eq{hr} (the normalization of the expectation values in the path integral is ensured by imposing $Z[0,0]=1$). Finally it is important to emphasize that the exponential of the full Hamiltonian must be treated {\it exactly} for the projection onto the ground state to work out. Such a treatment is not possible to carry out in interacting field theories with few exceptions. Therefore, one must be careful about the projection interpretation in perturbation theory.

\subsection{Perturbation Theory}

Although our discussion in this section so far forms the basis of the $i\e$ prescription in the operator formalism, it is also mostly abstract and not suitable for practical calculations. In this subsection we would like to formulate the in-in perturbation theory with $i\e$ terms, i.e. we will try to evaluate \eq{tvv2} perturbatively by treating the $i\e$ terms cautiously. 

As usual, perturbation theory can be most suitably formulated in the interaction picture as in subsection \ref{2b}. Defining the free deformed (Schr\"{o}dinger picture) evolution operator 
\be\label{uhe0}
U_0^\e(t)=T\exp\left[-i(1-i\e)\int_{-\infty}^t H_0(t')dt'\right],
\ee
one may insert identity operators in the numerator of \eq{tvv2} as  
\be\label{99}
\left<\psi|\right. U_H^\e{}^\dagger\, (U_0^\e{}^\dagger)^{-1}\,U_0^\e{}^\dagger\,O\,U_0^\e\,(U_0^\e)^{-1}U_H^\e\left.|\psi\right>.
\ee
One must note the distinct structure of the identity operators introduced in \eq{99} since  $U_0^\e$ is no longer unitary. The deformed interaction picture evolution operator can be defined as 
\be\label{uie}
U_I^\e(t)=U_0^\e(t)^{-1}U_H^\e(t)
\ee
and the numerator can be rewritten as
\be\label{n2}
\left<\psi|\right. U_I^\e(t)^\dagger\,O_I^\e(t)\,U_I^\e(t)\left.|\psi\right>,
\ee
where 
\be\label{oi2}
O_I^\e(t)=U_0^\e(t)^\dagger\,O\,U_0^\e(t).
\ee
It is possible to re-express $U_I^\e(t)$ in \eq{uie} for perturbation theory by first calculating its time derivative and then writing down the solution as 
\be\label{p1}
U_I^\e(t)=T\exp\left[-i(1-i\e)\int_{-\infty}^t H^+(t')\,dt'\right],
\ee
where 
\be\label{hpl}
H_I^+(t)=U_0^\e(t)^{-1}(H-H_0)U_0^\e(t).
\ee
The new definitions \eq{uie} and \eq{p1} must be compared to the earlier one \eq{ue2}. The conjugate of $U_I^\e$ can be found as 
\be\label{p2}
U_I^\e(t)^\dagger=\overline{T}\exp\left[i(1+i\e)\int_{-\infty}^t H^-(t')\,dt'\right],
\ee
where 
\be\label{hmi}
H_I^-(t)=H_I^+(t)^\dagger=U_0^\e(t)^{\dagger}(H-H_0)(U_0^\e(t)^\dagger)^{-1}.
\ee
This whole procedure is nearly identical to the transformations reviewed in subsection \ref{2b} except there are some minor complications of non-unitarity induced by the $i\e$ deformation. One should acknowledge that both $H$ and $H_0$ above are Schr\"{o}dinger picture Hamiltonians which are functions of the time independent Schr\"{o}dinger picture variables, although we do not explicitly indicate this as in subsection \ref{2b}. The denominator in \eq{tvv2} can also be rewritten as
\be \label{denom}
\left<\psi|\right. U_H^\e(t)^\dagger U_H^\e(t)\left.|\psi\right>=\left<\psi|\right. U_I^\e(t)^\dagger U_0^\e(t)^\dagger\,U_0^\e(t) U_I^\e(t)\left.|\psi\right>,
\ee
where we have used \eq{uie}. 

We are  ready to evaluate \eq{n2} and \eq{denom} perturbatively by expanding the exponentials in \eq{p1} and \eq{p2}. It is clear from our discussion in subsection \ref{gc} that unless one can sum the whole infinite series non-perturbatively, the projection from an arbitrary state onto the exact vacuum cannot happen.\footnote{Observe that the identity $\lim_{x\to\infty} e^{-x}c=0$ cannot be satisfied when the exponential is expanded in power series up to a finite order, which illustrates in an elementary example why the projection fails in perturbation theory.} However, projection onto the free ground state still takes place as follows: Imagine \eq{p1} is expanded up to a certain order in \eq{n2}. Since these are all time ordered, the free deformed evolution operators, with one exception, always appear with pairs in the form
\be\label{upa}
U_0^\e(t_1)U_0^\e(t_2)^{-1},\hs{5}t_1\geq t_2.
\ee
Note that the same is also true for the $U_0^\e(t)$  coming from \eq{oi2} because $t$ is the largest time. It is not difficult to observe that 
\be\label{upas}
U_0^\e(t_1)U_0^\e(t_2)^{-1}=T \exp\left[-i(1-i\e)\int_{t_2}^{t_1}\, H_0(t')\,dt'\right],\hs{5}t_1\geq t_2.
\ee
Unless $t_2\to-\infty$, which can only happen as a lower limit coming from the integral in \eq{p1}, one can set $\e=0$ in \eq{upas} (apparently, the simple role of the $\e$ terms is to make the integrals convergent near past infinity as we have anticipated). Besides, there is always one rightmost $U_0^\e(t')$ acting on $\left.|\psi\right>$ projecting it onto $\left.\vv$. The same argument can be repeated for the free deformed propagators acting to the left of $O$ in \eq{n2} and in the denominator \eq{denom}, which demonstrates that all states in \eq{tvv2} are actually projected onto the free vacuum (note that the unknown normalization factor that shows up in the projection, e.g. $c(\psi)$ in \eq{uevac}, cancels each other). Earlier, at the end of subsection \ref{3b}, we have pointed out that the free $\left.\vv$ and exact $\left.\tvv$ vacua are expected to be the same at past infinity for the self-interacting scalar field we are dealing with because the interaction potential is multiplied by $a^3$, which is the most diminishing factor in the Hamiltonian. Nevertheless, we prefer to distinguish them below keeping in mind more general situations.  

Let us now work out the $+$ branch evolution in more detail. The interaction Hamiltonian $H_I^+$, which is defined in \eq{hpl}, can be found for the scalar field as
\be
H_I^+=a^3\int d^3 x \,V(\f^+),
\ee 
where 
\be\label{fpev}
\f^+(t,\vx)=U_0^\e(t)^{-1}\,\f(\vx)\,U_0^\e(t).
\ee
One can also evolve the initial momentum $P_\f(\vx)$ by the same similarity transformation \eq{fpev} to obtain $P^+$, i.e. the momentum conjugate to $\f^+$. These satisfy the equal time commutation relation $[\f^+(t,\vx),P^+(t,\vec{y})]=i\d^3(\vx-\vec{y})$. The equations of motion that follow from \eq{fpev} can be determined by taking its time derivatives. It turns out that the first order derivative implies $\dot{\f}^+=(1-i\e)P^+/a^3$ and the equation involving the second order time derivatives can be solved by the mode expansion  
\be\label{fap}
\f^+(t,\vx)=\int \fr{d^3 k}{(2\pi)^{3/2}}\left[ e^{i\vc.\vx}\f^{+(1)}_k(t)\,\ak+e^{-i\vc.\vx}\f^{+(2)}_k(t)\,\akd\right],
\ee
where the mode functions are the two solutions of  
\be\label{fpl}
\ddot{\f}^+_k+3H\dot{\f}^+_k+(1-i\e)^2\left(m^2+\fr{k^2}{a^2}\right)\f_k^+=0
\ee
whose Wronskian is normalized $\f^{+(1)}_k\dot{\f}^{+(2)}_k-\dot{\f}^{+(1)}_k \f^{+(2)}_k=i(1-i\e)/a^3$ so that the canonical commutation relation is satisfied. The exact solutions of \eq{fpl} can be found by substituting $k\to(1-i\e)k$ and $m\to(1-i\e)m$ in the mode functions of the Bunch-Davies vacuum in de Sitter space. Since one demands to get the canonical field \eq{fa} as $\e\to0$, the mode functions $\f^{+(1)}_k$ and $\f^{+(2)}_k$ must be obtained from $\f_k^{BD}$ and $\f_k^{BD}{}^*$ of \eq{bdm} as
\be \label{119}
\f^{+(1)}_k=\f^{BD}_{k(1-i\e)},\hs{5}\f^{+(2)}_k=\f^{BD*}_{k(1-i\e)}.
\ee
Here we only indicate $k\to(1-i\e)k$ substitution, which yields the main modification of the asymptotic behavior. The $-$ branch can be obtained by noting that $\f^-(t,\vx)=\f^+(t,\vx)^\dagger$ and $H_I^-=a^3\int d^3 x \,V(\f^-)$. The careful reader would recognize  this construction as the operator equivalent of the path integral derivation discussed below \eq{hr}. It is easy to check that the asymptotic properties of the propagators listed in \eq{et2} are also valid here, which we will refer in a moment.  

In this perturbative series, the $+$ fields always appear to the right of $-$ ones and they are time and anti-time ordered, respectively (we will show below that $O^\e_I(t)$ can be expressed in terms of $+$ fields). These can be evaluated by using Wick's theorem in terms  of the propagators \eq{cf}, and \eq{et2} can be used as $\e\to0$. In that limit, $\f^\pm(t,\vx)\to\f(t,\vx)$ for any finite time $t$ and the $\e$ deformation only shows up in \eq{et2}. Since these $\e$ terms simply make otherwise divergent time integrals near infinity to be convergent, they can be replaced with a single factor  in \eq{et2} doing the same job and yielding identical results as $\e\to0$ while erasing the difference between $\pm$ branches. Thus, one can substitute {\it for all times} 
\be
\f^\pm(t,\vx)\to\f^\e(t,\vx),
\ee
where the field $\f^\e(t,\vx)$ is obtained from $\f(t,\vx)$ in \eq{fa} by modifying the mode functions with a multiplicative factor, say $e^{\e\, k \,\eta}$, as follows
\be\label{fep}
\f^\e(t,\vx)=\int \fr{d^3 k}{(2\pi)^{3/2}}\left[ e^{i\vc.\vx}\f_k^{BD}(t)e^{\e\, k \,\eta}\,\ak+e^{-i\vc.\vx}\f^{BD*}_k(t)e^{\e\, k \,\eta}\,\akd\right].
\ee
As it turns out such a convergence term is also necessary to get a well defined position space de Sitter propagator, see e.g \cite{i1}. We also note that if this modification had been utilized in \eq{hbd}, the second unwanted line would vanish in the limit which makes $\left.\vv$ a proper eigenstate of the free Hamiltonian at past infinity. 

Another source of asymmetry between $\pm$ branches in the perturbative series is the factors $(1\mp i\e)$ multiplying $H_I^\pm$ coming from \eq{p1} and \eq{p2}. However, these can be set to unity in the limit $\e\to0$ when the exponentials are expanded in power series up to a certain order since they multiply necessarily finite quantities in the same limit (other possibility indicates a problem in the theory). 

For any Schr\"{o}dinger picture (polynomial) operator, like $O=\f(\vx_1)\f(\vx_2)$, one may obtain by using the definitions \eq{oi2} and \eq{fpev} that 
\be
O_I^\e(t)= U_0^\e(t)^\dagger\,U_0^\e(t)\,O_I^+(t)\to U_0^\e(t)^\dagger\,U_0^\e(t)\,O_I(t),
\ee
where $\f^+$ fields are replaced by $\f$ since $t$ refers to a finite time and $O_I(t)$ is the standard interaction picture operator which can be obtained by using \eq{fa}.

Combining all these findings, one may see that \eq{tvv2} can be rewritten as
\be\label{v0}
\vvc\right. O\left.\vv(t)=\lim_{\e\to0}\,\fr{\vvc \right.U_I^\dagger[\f^\e]\, U_0^\e(t)^\dagger\,U_0^\e(t)\,O_I(t)\, U_I[\f^\e]\left.\vv}{\vvc\right. U_I^\dagger[\f^\e] U_0^\e(t)^\dagger\,U_0^\e(t)\,U_I[\f^\e]\left.\vv},
\ee
where $U_I[\f^\e]$ is the standard interaction picture {\it unitary} time evolution operator \eq{fui} which is expressed in terms of the Hermitian field $\f^\e$ introduced in \eq{fep} and we have also worked out the denominator along the same lines. As pointed out below \eq{upas}, the unique effect of $\e$ terms in \eq{n2} and \eq{denom} was to make the time integrals convergent at past infinity and to project onto the free vacuum. In \eq{v0}, the convergence is achieved by the field $\f^\e$ and the projection is already taken into account, thus one can safely set $\lim_{\e \to 0}U_0^\e(t)^\dagger\,U_0^\e(t)=I$, which finally gives 
\be\label{vf}
\vvc\right. O\left.\vv(t)=\lim_{\e\to0}\,\vvc \right.U_I^\dagger[\f^\e]\, O_I(t)\, U_I[\f^\e]\left.\vv.
\ee
Note that the normalization is evident since $U_I[\f^\e]$ is a unitary operator and choosing $O=I$ gives $O_I(t)=I$. One may expand \eq{vf} perturbatively in powers of the potential to get Weinberg's commutator formula \eq{inp} where all integrals near minus infinity are simply made convergent by the damping $\e$ terms in $\f^\e$. 

\subsection{Comparison to other Prescriptions and Absence of Spurious Divergences}

It is interesting to compare our final result \eq{vf} with the standard methods used in the literature. In particular, we have discussed that one usually employs the $i\e$ prescription by defining the complexified unitary evolution operator \eq{ue2}, which is then used in \eq{tvv}. In \cite{c3}, it is pointed out that instead of deforming the time integration contour into the complex plane as in \eq{ue2}, one may instead utilize the replacement\footnote{The mode functions used in \cite{c3} has the opposite sign for the conformal time, i.e. one has $\eta\to-\eta$ as compared to the standard Bunch-Davies mode functions, see e.g. \cite{wein1}. Thus their $i\e$ prescription also differs by a sign.}   
\be\label{tr1}
H_I(t)\to H_I(t(1-i\e)),
\ee
where $t$ is real and the integration in $U_I$ is carried over the real line (it is not difficult to see that both methods \eq{ue2} and \eq{tr1} are indeed equivalent). Obviously, there is a subtlety in applying this prescription in the presence of  non-analytical time dependence and one should introduce appropriate cuts in the complex plane. Fortunately in de Sitter space (or in a slow-roll inflationary background) this is not a problem for the Bunch-Davies vacuum, which is the main subject of interest. 

For a self-interacting scalar field with a polynomial potential, \eq{tr1} is equivalent to
\be\label{tr2}
\f_I(t,\vx)\to \f_I(t(1-i\e),\vx).
\ee
Let us consider a massless scalar in de Sitter space and try to understand the consequences of \eq{tr2} in the in-in perturbation theory. In such a calculation one encounters the following four different types of Wick contractions 
\bea
&&\vvc\right. T\f_I(t_1(1-i\e))\f_I(t_2(1-i\e)) \left.\vv,\hs{5}\vvc\right. \f_I(t_*)\f_I(t_1(1-i\e)) \left.\vv,\nn\\
&&\vvc\right. \overline{T}\f_I(t_1(1+i\e))\f_I(t_2(1+i\e)) \left.\vv,\hs{5}\vvc\right. \f_I(t_1(1+i\e))\f_I(t_*) \left.\vv,\label{wy2}
\eea
where $\f_I(t_*)$ comes from $O_I(t_*)$, $t_*\geq t_1,t_2$ and we suppress the spatial dependence of the fields. The interaction picture field can be expanded in terms of the massless Bunch-Davies mode function 
\be\label{fai}
\f_I(t,\vx)=\int \fr{d^3 k}{(2\pi)^{3/2}}\fr{1}{a\sqrt{2k}}\left[ e^{i\vc.\vx-ik\eta}\,\left[1-\fr{i}{k\eta}\right]\,\ak+e^{-i\vc.\vx+ik\eta}\,\left[1+\fr{i}{k\eta}\right]\,\,\akd\right]
\ee
and this expansion can be used in \eq{wy2}. A shift in the proper time  $t\to t(1\pm i\tilde{\e})$ implies a similar shift in the corresponding conformal time $\eta\to\eta(1\pm i\e)$ and one may see that all contractions in \eq{wy2} include suitable convergence factors
\be\label{conv1}
e^{-\e\,k \,\Delta\eta}
\ee
where $k$ is the momentum of the corresponding two point function as defined in the footnote \ref{foot6} and $\Delta\eta\geq0$ due to time orderings so that $\Delta\eta\to+\infty$ when $\eta_1\to-\infty$ or $\eta_2\to-\infty$ (for example $\Delta \eta=-\eta_1$ in the last two contractions or $\Delta \eta=(\eta_1-\eta_2)\th(\eta_1-\eta_2)+(\eta_2-\eta_1)\th(\eta_2-\eta_1)$ for the time ordered Green function). One may worry that the convergence factor may fail when $\eta_1$ and $\eta_2$ simultaneously approach to $-\infty$, however no problem arises in explicit calculations presumably because the line $\eta_1=\eta_2$ is of measure zero in the double integral  $\int d\eta_1 d\eta_2$. 

Let us now similarly examine the expression \eq{vf} in the in-in perturbation theory. As \eq{vf} can be written in the form of nested commutators, it can also be evaluated by using Wick contractions like in \eq{wy2}, where now $\f^\e(t,\vx)$ in \eq{fep} is used for the interaction picture field and no $i\e$ shift of the time arguments is applied. It is easy to see that, these contractions have one of the following terms 
\be\label{conv2}
e^{\e\,k \,\eta_1},\hs{10}e^{\e\,k \,\eta_2},\hs{10}e^{\e k(\eta_1+\eta_2)},
\ee
which guarantee convergence as $\eta_1\to-\infty$ and/or $\eta_2\to-\infty$. Therefore, both prescriptions provide similar asymptotic behavior \eq{conv1} and \eq{conv2} that can cure the oscillating non-convergent time integrals near past infinity. 

\begin{figure}
\centerline{\includegraphics[width=9cm]{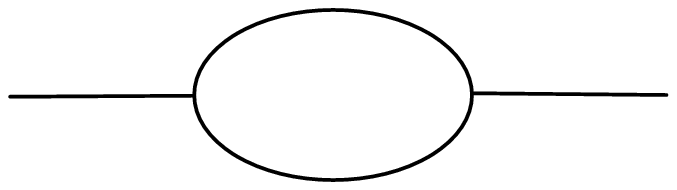}}
\caption{A one loop graph in $g\f^3$ theory that may contain a spurious divergence if the $i\e$ prescription is not utilized properly.} 
\label{fig2}
\end{figure}

As discussed in \cite{c3}, the naive commutator formula \eq{inp} requires a tricky usage of the $i\e$ prescription that otherwise possibly incorporates spurious divergences. Our final formula \eq{vf}, which can also be expressed as nested commutators, is free from this issue since we have just shown that it implements the same convergence with the analytical continuation method. Still, to illustrate the absence of spurious divergences explicitly in a loop effect let us consider the second order term in \eq{inp}, which can also be expressed as   
\be\label{131}
\left< O(t_*)\right>^{(2)}=-2\, {\cal R}e\,\int_{-\infty}^{t_*} dt_2\,
\int_{-\infty}^{t_2} dt_1 \, \left[\left<H_I(t_1)H_I(t_2)O_I(t_*)\right>-\left<H_I(t_1)O_I(t_*)H_I(t_2)\right>\right].
\ee
Let us take the two-point function $O(t_*)=\f(t_*,\vec{k})\,\f(t_*,\vec{l})$ and calculate the correction that arises from a cubic potential $V=g\f^3$ giving the interaction Hamiltonian
\be
H_I(t)=g\,a^3(t)\,\int d^3 x\, \f_I^3(t,\vx). 
\ee
The graph of the connected contribution can be pictured as in Fig. \ref{fig2}. Using the mode expansion of fields, a straightforward calculation gives
\bea
-\fr{72g^2}{(2\pi)^3}\,&&\d^3\,(\vec{k}+\vec{l})\,{\cal R}e\,\int d^3p\,\int_{-\infty}^{t_*} dt_2\,\int_{-\infty}^{t_2} dt_1\, a^3(t_2)\,a^3(t_1)\label{133}\\
&&\f^*_p(t_2)\,\f^*_{|p-k|}(t_2)\,\f_p(t_1)\,\f_{|p-k|}(t_1)\,\left[\f_k(t_2)\,\f_k(t_1)\,\f_k^2(t_*)-\f_k^*(t_2)\,\f_k(t_1)\,|\f_k(t_*)|^2\right],\nn
\eea
where the terms in the square brackets in \eq{133} come correspondingly from the ones in \eq{131}. Here, a spurious divergence may emerge from the second term in the square brackets in \eq{133}. Indeed, expressing the integrals in conformal time and using the massless mode functions, this second term can be seen to have the following form 
\be
\int_{-\infty}^{\eta_*}\, \fr{d\eta_2}{\eta_2^4}\,e^{i(k+p+|p-k|)\eta_2}\left[...\right]\int_{-\infty}^{\eta_2}\,\fr{d\eta_1}{\eta_1^4}\,e^{-i(k+p+|p-k|)\eta_1}\left[...\right]
\ee 
where the dotted terms involve cubic order polynomials of conformal times. One may note that the phase coming from the upper limit of the $\eta_1$ integration exactly cancels out the phase in the $\eta_2$ integral, which then becomes spuriously divergent. 

In our way of utilizing the $i\e$ prescription, this possible divergence is avoided naturally. From \eq{fep}, the mode functions are given by 
\be
\f_k(t)=\f_k^{BD}(t)e^{\e\, k \,\eta},
\ee
which ensures the convergence of {\it both} $\eta_1$ and $\eta_2$ integrals. Taking the real part $2\,{\cal R}e\,z=z+z^*$ and relabeling the integration variables $\eta_1\leftrightarrow\eta_2$ in $z^*$, the potentially problematic term in \eq{133} becomes
\be\label{136}
\fr{36g^2}{(2\pi)^3}\,\d^3\,(\vec{k}+\vec{l})\,|\f_k(t_*)|^2\int d^3 p\left| \int_{-\infty}^{t_*} dt\, a^3(t)\,\f_k(t)\,\f_p(t)\,\f_{|p-k|}(t)\right|^2.
\ee
This contains a nonzero oscillating phase and $\e$ can safely be set to zero after carrying out the time integral without any divergence. One may find that the analytical continuation method also yields the same expression \eq{136}, see e.g. \cite{c3} that worked out a similar correction. 

\section{Conclusions}

In this paper we try to formalize the $i\e$ prescription in cosmology for a self-interacting scalar field propagating in the Poincare patch of the de Sitter space. The problem is completely different than the flat space counterpart due to  reasons like the presence of an explicit time dependence and the absence of an invariant ground state. Besides, one would like to utilize the prescription in the in-in theory for the expectation values rather than in the usual in-out theory dealing with the scattering amplitudes. The basic need for the $i\e$ prescription in cosmology stems from the divergent oscillating time integrals near past infinity encountered in perturbation theory, which are simply cured by giving a small imaginary component to the time variable like in the flat space prescription. While the naive application of the flat space reasoning can be justified on the basis that at very early times all the modes of interest are subhorizon and in the Bunch-Davies vacuum they must mimic the flat space behavior, there are still some issues in this procedure and also a derivation from first principles is lacking. Especially, it is not clear at all how one can justify the standard interpretation that the $i\e$ prescription projects onto the exact ground state because of the explicit time dependence in the problem. 

We observe that the real issue is connected to the infinite temporal unitary evolution which produces ambiguous infinite phases. In the free theory, this does not create any problem when one calculates the vacuum expectation values since the phases cancel each other. We have seen, however, that the indefinite phases can be problematic even in the free theory when the expectation values in the asymptotically superposed states are considered. In the presence of interactions, it is precisely these infinite phases that give rise to oscillating divergent integrals in the perturbation theory.   

The $i\e$ prescription can be viewed as a technique that resolves the indefinite phase issue related to the temporal asymptotics by introducing some form of non-unitarity in a controlled way. In the path integral approach, often neglected vacuum wave-functionals suggest how to incorporate the required non-unitarity. Like in the flat space in-out path integrals, the vacuum wave-functionals of the free theory induce $i\e$ terms in the quadratic in-in path integral which modify the temporal asymptotic fall off behavior of the in-in propagators by $\e$ dependent decreasing factors. These new terms not only cure the divergence problem of the time integrals in the perturbation theory but they also project onto the free vacuum in the expectation values.  
 
The path integral method guides on how to incorporate the $i\e$ prescription in the operator formalism by altering the time evolution by the scaled Hamiltonian $(1-i\e)H$. One must appreciate that this deformation is completely different than complexifying the time variable which becomes intricate when the explicit time dependence is possibly non-analytic. Not surprisingly, the modification asymptotically kills all non-vacuum components of a given initial state and projects onto the vacuum at past infinity. We have demonstrated this property in the free theory by explicit calculations and gave a heuristic argument in general, which are consistent with the path integral considerations. Moreover, we have shown that the proposed $i\e$ prescription also yields the appropriate decreasing factors that make the time integrals in the perturbation theory convergent at past infinity. 

We have seen that for explicit calculations the final outcome of using the $i\e$ prescription is to introduce appropriate convergence factors in the time integrals arising either in Weinberg's commutator formula or in the diagrammatic expansions. Thus, the present work offers a first principles derivation of the commonly used technique in the literature. Moreover, our treatment also addresses some concerns about utilizing the $i\e$ prescription in Weinberg's commutator formula. Finally, there is an evident puzzle in the standard statement that the $i\e$ prescription projects onto the interacting vacuum of the theory: Can't we just then get some nontrivial information about the exact ground state by using the $i\e$ prescription? As we have discussed, that statement unfortunately fails in the perturbation theory where only a projection onto the free vacuum occurs. Any investigation about the exact ground state requires some form of non-perturbative treatment. As always, there is no free lunch and the $i\e$ prescription cannot give any useful knowledge about the exact vacuum. 
 
\begin{acknowledgments}
	
This work has been done during author's mandatory sabbatical at home. I would like to thank my wife Y\i ld\i z for her support and patience, my sons Mahir and Zafer for discussions! on diverse elementary school topics and my son Sami for  various Skype conversations. I also thank Richard Woodard for a useful e-mail correspondence. 

\end{acknowledgments}

\end{document}